\documentclass[sigconf,nonacm]{acmart}
\newif\ifcameraready
\camerareadytrue

\newcommand{\OpenHandsArtifactURL}{https://github.com/vals-ai/VibeCodeBench-Openhands-Scaffold}
\newcommand{\VCBTrajectoriesURL}{https://github.com/vals-ai/vibe-code-bench-paper-artifacts}

\usepackage{listings}
\usepackage{subcaption}

\begin{document}

\title{Vibe Code Bench: Evaluating AI Models on End-to-End Web Application Development}

\author{Hung Tran}
\email{hung@vals.ai}
\affiliation{%
  \institution{Vals AI}
  \city{San Francisco}
  \state{California}
  \country{USA}
}

\author{Langston Nashold}
\email{langston@vals.ai}
\affiliation{%
  \institution{Vals AI}
  \city{San Francisco}
  \state{California}
  \country{USA}
}

\author{Rayan Krishnan}
\email{rayan@vals.ai}
\affiliation{%
  \institution{Vals AI}
  \city{San Francisco}
  \state{California}
  \country{USA}
}

\author{Antoine Bigeard}
\email{antoine.bigeard.cs@gmail.com}
\affiliation{%
  \institution{Vals AI}
  \city{San Francisco}
  \state{California}
  \country{USA}
}

\author{Alex Gu}
\email{gua@mit.edu}
\affiliation{%
  \institution{Massachusetts Institute of Technology}
  \city{Cambridge}
  \state{Massachusetts}
  \country{USA}
}

\begin{abstract}
Code generation has emerged as one of AI's highest-impact use cases, yet existing benchmarks measure isolated tasks rather than the complete "zero-to-one" process of building a working application from scratch. We introduce \textbf{Vibe Code Bench}, a benchmark of 100 web application specifications (50 private validation, 50 held-out test) with 964 browser-based workflows comprising 10,131 substeps, evaluated against deployed applications by an autonomous browser agent.

Across 16 frontier models, the best achieves 61.8\% accuracy on the test split, revealing that reliable end-to-end application development remains a frontier challenge. We identify self-testing during generation as a strong performance predictor (Pearson r=0.72), and show through a completed human alignment study that evaluator selection materially affects outcomes (31.8–93.6\% pairwise step-level agreement).

Our contributions include (1) a novel benchmark dataset and browser-based evaluation pipeline for end-to-end web application development, (2) a comprehensive evaluation of 16 frontier models with cost, latency, and error analysis, and (3) an evaluator alignment protocol with both cross-model and human annotation results.

We provide reproducibility artifacts, detailed in Appendix \ref{sec:appendix-artifacts}.
\end{abstract}

\ccsdesc[500]{Computing methodologies~Natural language processing}
\ccsdesc[500]{Software and its engineering~Software testing and debugging}
\ccsdesc[500]{Software and its engineering~Software development techniques}
\ccsdesc[300]{Computing methodologies~Neural networks}

\keywords{code generation, benchmarks, large language models, software engineering, automated evaluation, agentic coding, web development}

\maketitle

\section{Introduction}

\paragraph{Motivation}
Building software is a high-leverage way to translate ideas into economic value, but the ability to implement those ideas remains concentrated among a small fraction of people. Meanwhile, code generation has become a major use case for generative AI, and developer-facing tools are seeing rapid adoption in day-to-day workflows~\cite{peng2023copilot, stackoverflow2024survey}. Existing benchmarks suggest that models are increasingly capable on constrained tasks (e.g., function synthesis and issue resolution)~\cite{chen2021codex, austin2021program, jimenez2024swebench, jain2024livecodebench}.

However, the most consequential promise of ``vibe coding'' is not faster completion of isolated edits; it is enabling non-technical users to build complete applications end-to-end from natural language intent. If reliable, this capability could expand who can create software and compress the time from specification to a working prototype.

\paragraph{The gap.}
Despite extensive work on code generation evaluation, there is no widely used benchmark that measures whether models can produce a complete, deployable web application from a natural language specification---a ``zero-to-one'' setting that includes multi-file code, configuration, deployment, and user-facing workflows.

\paragraph{Overview of Vibe Code Bench.}
We introduce \textbf{Vibe Code Bench (VCB)}, the first benchmark to test LLMs' ability to generate a complete working web application from only a text specification. It contains 100 realistic application specifications split into a \textbf{private validation set} (50 tasks) and a \textbf{test set} (50 tasks). Across both splits, tasks are paired with 964 automated browser workflows (10,131 substeps). Each task asks a model to build an application from scratch in a sandboxed environment, with access to a browser, a terminal, and common production services (e.g., authentication, databases, payments, and email). Applications are evaluated by an autonomous browser agent that executes end-to-end workflows and scores success based on substep completion.

\begin{figure*}[!t]
\centering
\includegraphics[width=0.90\textwidth]{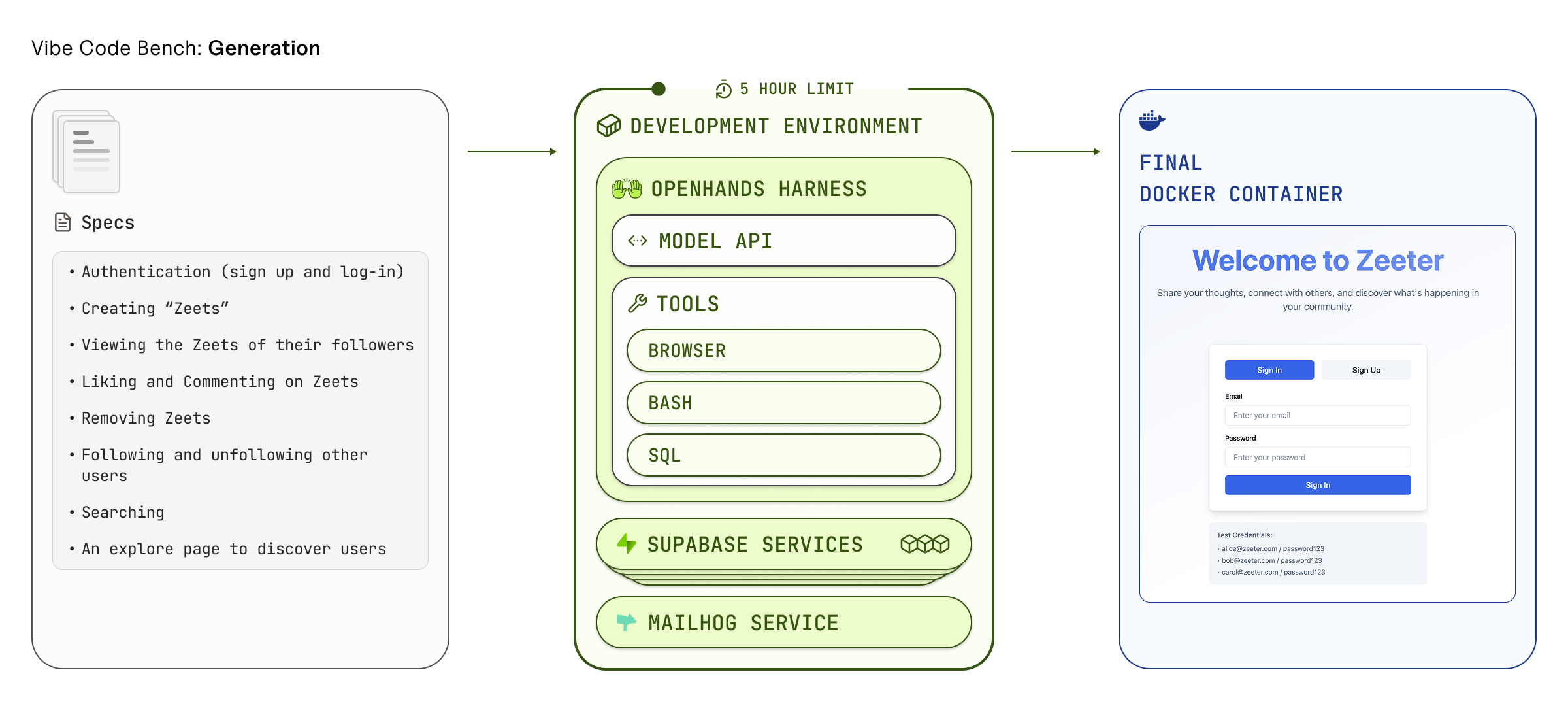}
\caption{Generation flow from natural-language specification to a runnable application artifact.}
\label{fig:pipeline}
\Description{Generation-flow schematic showing the benchmark's app construction pipeline from input specification through agent execution to final runnable artifact.}
\end{figure*}

\paragraph{Key findings.}
Across 16 frontier models, we find that end-to-end application development remains unreliable: the top model (GPT-5.3-Codex) passes 61.8\% of workflows. We also observe a strong association between a model's self-testing behavior (browser usage during development) and its final performance. Our alignment study shows that evaluator choice matters: some evaluator pairs align strongly, whereas others diverge substantially.

\paragraph{Contributions.} Our contributions include the following:

\begin{enumerate}
    \item \textbf{A novel benchmark dataset} of 100 natural-language web application specifications split into a private validation set and a test set, with 964 browser-based workflows spanning diverse domains and difficulty levels.
    \item \textbf{A realistic generation harness} (based on a fork of OpenHands ~\cite{wang2024openhands}) that supports multi-hour development sessions with browser and terminal access in a containerized environment.
    \item \textbf{An automated evaluation pipeline} using browser agents to test end-to-end workflows and produce fine-grained pass-fail signals.
    \item \textbf{A comprehensive evaluation} of 16 frontier models across providers, including cost/latency tradeoffs and error analyses. Performance on the Vibe Code Bench is far more discriminative than on existing benchmarks like SWE-Bench: the gap between MiniMax M2.5 and Claude Opus 4.6 is 2.8\% on SWE-Bench ~\cite{swebenchleaderboard2026} but 42.7\% on Vibe Code Bench.
    \item \textbf{An evaluator-alignment protocol and analysis}, including completed cross-evaluator robustness experiments and an external three-annotator human-alignment study.
\end{enumerate}
\begin{table}[h]
\caption{Top-model accuracy on Vibe Code Bench (test split).}
\label{tab:headline_accuracy}
\small
\begin{tabular}{@{}lr@{}}
\toprule
\textbf{Model} & \textbf{Accuracy $\pm$ SE (\%)} \\
\midrule
GPT-5.3-Codex & $61.77 \pm 4.71$ \\
Claude Opus 4.6 & $57.57 \pm 4.37$ \\
GPT-5.2 & $53.50 \pm 5.07$ \\
Claude Opus 4.6 Thinking & $53.50 \pm 4.68$ \\
Claude Sonnet 4.6 & $51.48 \pm 4.64$ \\
GPT-5.2-Codex & $37.91 \pm 4.58$ \\
\bottomrule
\end{tabular}
\end{table}

\section{Related Work}
We situate Vibe Code Bench within the broader landscape of code generation benchmarks, web agent evaluation, and agentic coding systems.

\paragraph{Code Generation Benchmarks.}
Existing benchmarks span the spectrum from function-level to repository-level tasks, but none evaluate the full ``zero-to-one'' application development process. HumanEval~\cite{chen2021codex} and MBPP~\cite{austin2021program} evaluate function-level code generation from docstrings, testing isolated algorithmic reasoning. SWE-Bench~\cite{jimenez2024swebench} and its variants (SWE-Bench Verified~\cite{chowdhury2024swebenchverified}, SWE-Bench Pro~\cite{swebenchpro2025}) test models' ability to resolve real GitHub issues within existing codebases, requiring an understanding of complex codebases but assuming substantial existing infrastructure. SWE-Lancer~\cite{miserendino2025swelancer} evaluates models on \$1M worth of real freelance tasks from Upwork, providing economic grounding but still focusing on scoped modifications rather than greenfield development. LiveCodeBench~\cite{jain2024livecodebench} provides continuously updated competitive programming problems to avoid contamination.

Table~\ref{tab:benchmark_comparison} summarizes key differences between existing benchmarks and Vibe Code Bench. The critical gap VCB addresses is the \textbf{zero-to-one} capability: building complete, deployable applications from natural language specifications rather than modifying existing code or solving isolated problems.

\begin{table}[t]
\caption{Comparison of code-generation benchmarks.}
\label{tab:benchmark_comparison}
\small
\setlength{\tabcolsep}{4pt}
\begin{tabular}{@{}p{2.0cm}p{1.5cm}p{1.7cm}p{2.5cm}@{}}
\toprule
\textbf{Benchmark} & \textbf{Task Type} & \textbf{Eval Method} & \textbf{Gaps} \\
\midrule
HumanEval & Single function & Unit tests & No multi-file apps or deployment \\
SWE-Bench & Bug fixes & Unit Tests & Assumes existing codebase \\
SWE-Lancer & Freelancing & E2E tests & Scoped tasks, not full apps \\
LiveCodeBench & Leetcode & Unit tests & No real-world context \\
Terminal-Bench & Terminal Tasks & Unit Tests & No browser UI or persistence \\
WebArena & Web navigation & Task Success & Navigation only, no code gen \\
\midrule
\textbf{VCB} & \textbf{Full app} & \textbf{Browser agent} & \textbf{---} \\
\bottomrule
\end{tabular}
\end{table}

\paragraph{Web Agent Benchmarks.}
WebArena~\cite{zhou2024webarena} and VisualWebArena~\cite{koh2024visualwebarena} evaluate agents on web navigation and interaction tasks, achieving up to 60\% success rates on complex multi-step workflows. These benchmarks test the ability to \emph{use} existing web applications but not to \emph{build} them. Our evaluation pipeline leverages similar browser automation techniques but applies them to test AI-generated applications rather than human-built ones.

\paragraph{Agentic Coding Systems.}
Academic work has developed open-source frameworks for extended coding sessions. OpenHands~\cite{wang2024openhands} provides a platform for AI software developers with browser and terminal access. SWE-agent~\cite{yang2024sweagent} introduces agent-computer interfaces for automated software engineering, with mini-swe-agent offering a simplified 100-line variant. Agentless~\cite{xia2024agentless} takes a simpler approach without persistent agent state. Terminal-Bench~\cite{merrill2025terminalbench} evaluates agents on realistic command-line tasks.

Commercial systems---Claude Code~\cite{anthropic2024claudecode}, Cursor~\cite{cursor2024}, Codex~\cite{openai2025codex}, Replit Agent~\cite{replit2025agent}, and Lovable~\cite{lovable2025}---demonstrate that multi-hour development sessions are feasible in production. However, systematic evaluation of end-to-end application development capabilities has been limited. Vibe Code Bench provides a standardized framework for such evaluation.

\section{Benchmark Design}

Our benchmark design optimizes for three goals: (1) \textbf{realism} (tasks resemble the asks of non-technical users), (2) \textbf{reproducibility} (standardized environment and harness), and (3) \textbf{implementation-agnostic evaluation} (apps are judged by user-visible behavior rather than specific implementation).

\subsection{Data Construction}
\label{sec:data-construction}

Our benchmark consists of 100 application specifications designed to represent real-world software development scenarios, split into a \textbf{private validation set} (50 tasks) and a \textbf{test set} (50 tasks). The two splits are disjoint. Specifications span three domain categories:
\begin{itemize}
    \item \textbf{Individual} (24 apps): An app a user would vibe-code for their personal use. These most often do not have authentication. Example: an app to keep track of your habits.
    \item \textbf{Solo Founder} (45 apps): An app someone starting a business might build themselves. These often have simple email-based authentication, though some require more complex authentication. Example: an app to find and reserve parking in crowded areas.
    \item \textbf{Enterprise Tool} (31 apps): A program that a business may build internally rather than acquire from a vendor. These often include multiple roles or a role hierarchy. Example: an app to track and approve procurement requests.
\end{itemize}

We generated specifications and workflows as follows:
\begin{enumerate}
    \item \textbf{Idea generation}: Candidate tasks were inspired by consumer apps, YC startup ideas, consulting case studies, and common internal enterprise needs.
    \item \textbf{Draft generation}: A structured-output GPT-5 pipeline produced each initial specification and 6--23 draft UI workflows covering core functionality, edge cases, and critical user workflows.
    \item \textbf{Two-pass human review}: Two PM reviewers edited and finalized each task using a shared rubric, per-test Accept/Modify/Reject decisions, and a completion checklist.
    \item \textbf{Audit and freeze}: Before freeze, PMs line-reviewed each workflow to ensure every element tested was given in the specification. An LLM-based audit then flagged any residual spec/test misalignment, test steps not possible in the browser, anonymous-auth inconsistencies, and isolation violations between tests. Any final issues flagged were fixed by human revision.
\end{enumerate}

\paragraph{Task Format and Test Structure.}
Each application includes metadata (domain, difficulty, required services) and a natural-language specification shown to the models. The specification was designed to be concise and written from a non-technical perspective, to mimic the usage of a real vibe-coder.

Specifications must contain the information needed to pass the tests. The heuristic used was that any feature tested \textit{must} either be in the specification directly, or a qualified engineer would interpret it as clearly following from the specification. Any concrete emails, usernames, organization names, or form values must also be either in the spec or customizable by the testing agent. See Appendix~\ref{sec:appendix-spec-example} for an example.

\paragraph{Third-Party Service Integration.}
Across the benchmark, 28\% (28/100) of applications require integration with at least one external service (Stripe and/or email):
\begin{itemize}
    \item \textbf{Email only} (9 apps): User notifications, verification emails, reports
    \item \textbf{Stripe only} (6 apps): Payment processing, subscriptions, invoicing
    \item \textbf{Both email + Stripe} (13 apps): Workflows requiring cross-service consistency
\end{itemize}

These integrations are intentionally included for two reasons. First, they increase task difficulty in ways not captured by pure CRUD apps (e.g., async webhooks, delivery verification, and multi-service state consistency). Second, they better approximate real production requirements where payments and notifications are common. All services run in sandbox/test mode as isolated containers inside the generation environment.

\paragraph{Tests}
Applications are paired with UI workflows. The goal of these workflows is to evaluate performance as a real user would, clicking through the web application end-to-end in the browser.

\begin{itemize}
    \item \textbf{Task}: A task is a synonym for a single specification. An \textbf{application} is a single runnable implementation generated by the agent to solve a task.
    \item \textbf{Workflow}: Each task has between 6 and 23 workflows (or tests), designed to evaluate specific aspects of functionality. For example, a workflow for a social media platform involves creating an account, creating a new post, commenting on the post, and then signing out. Each workflow includes a brief purpose statement and a set of substeps.
    \item \textbf{Substep}: The individual steps the evaluator must take for each workflow are referred to as substeps. Examples include "Create an account with the email X" or "Navigate to Page Y and click Button Z".
\end{itemize}

These definitions are used by a browser-based agent to test the application (see Section~\ref{sec:automated-eval}). We chose a browser agent because it is implementation-agnostic: it avoids brittle DOM-coupled checks (e.g., fixed selectors) and does not require internal code instrumentation, in contrast to methods such as unit tests or programmatic end-to-end tests like Playwright. Models are free to choose UI structure and styling as long as user-visible behavior satisfies the specification. Appendix~\ref{sec:appendix-ui-test-example} provides an example set of workflow definitions.

\begin{table}[h]
\caption{Dataset splits and evaluation workload.}
\label{tab:splits}
\small
\begin{tabular}{@{}lrrr@{}}
\toprule
\textbf{Split} & \textbf{\#Tasks} & \textbf{\#Workflows} & \textbf{\#Substeps} \\
\midrule
Validation (public) & 50 & 491 & 4,995 \\
Test & 50 & 473 & 5,136 \\
\midrule
Total & 100 & 964 & 10,131 \\
\bottomrule
\end{tabular}
\end{table}

\subsection{Generation Harness}
\label{sec:generation-harness}

Models operate within a fully-featured development environment based on a modified version of OpenHands~\cite{wang2024openhands}, an agentic scaffold for programming tasks. We chose OpenHands for its robust containerization architecture, browser integration, extensible tool system, and wide adoption in the community.

\paragraph{Environment Components.}
\begin{itemize}
    \item \textbf{Isolated containers}: Each model receives an isolated container with complete control over the development workspace.
    \item \textbf{Terminal access}: Unrestricted access to install dependencies, run builds, start servers, and execute arbitrary commands.
    \item \textbf{Browser capabilities}: Web browsing for documentation lookup and self-testing generated applications.
    \item \textbf{Service integrations}: We provide Supabase, MailHog, and Stripe endpoints to the agent. The harness exposes connectivity and credentials, but the model must discover and use these services correctly in its generated app and workflows.
\end{itemize}

We use Supabase as the default backend substrate because it integrates cleanly with agent tooling while remaining straightforward to self-host reproducibly~\cite{supabase2024}. It unifies PostgreSQL, authentication, and object storage behind a single local stack, reducing setup variance across tasks.

\paragraph{MailHog and Stripe.}
We include MailHog and Stripe in the specifications and harness because many product workflows require outbound email and payment operations. Excluding these systems would overestimate real-world readiness by removing failure modes around external service wiring, retries, and state synchronization.

\paragraph{Available Tools.}
Agents have access to core OpenHands tools for terminal commands, file editing, web navigation, and task submission. The harness is able to access the full set of tools exposed by the Supabase and Tavily MCPs (Tavily is integrated with OpenHands by default). See Appendix \ref{sec:provided-tools} for the full set of tools.

\paragraph{Normalization.}
We normalize generation effort by giving every model the same wall-clock budget: up to \textbf{5 hours} per application. Any work completed within this window counts toward the final artifact; work after the timeout is not included. We use this time-based normalization because it mirrors real-world development constraints and it allows for the evaluation of a diverse set of harnesses, even those without instrumentation for counting turns.

\paragraph{System Prompt Design.}
We developed a detailed system prompt through iterative testing, addressing common failure modes: mandating a consistent tech stack (React + Vite frontend, Tailwind CSS, Supabase backend), adding explicit instructions to test deployment before submission, and providing detailed guidance on environment variable handling and Docker networking.

It was necessary to mandate a tech stack; otherwise, generated apps were too heterogeneous to evaluate consistently at scale. We also mandate Docker Compose so each submission is portable and can be started by the evaluator using a deterministic entry point.

\paragraph{Docker-in-Docker.}
Docker-in-Docker is required because agents must build and run app-specific Compose stacks (frontend, backend, and auxiliary services) inside isolated per-task workspaces. The outer container is the model sandbox; inner containers are application services launched by the agent. This design provides strong task isolation, reproducible startup behavior for evaluation, and portable artifacts that can be replayed later.

\subsection{Automated Evaluation Pipeline}
\label{sec:automated-eval}

Our evaluation pipeline uses Browser Use~\cite{browseruse2024}, an autonomous web agent, to perform point-and-click testing of generated applications. See Section~\ref{sec:data-construction} for test-definition structure.

\begin{figure*}[!t]
\centering
\includegraphics[width=\textwidth]{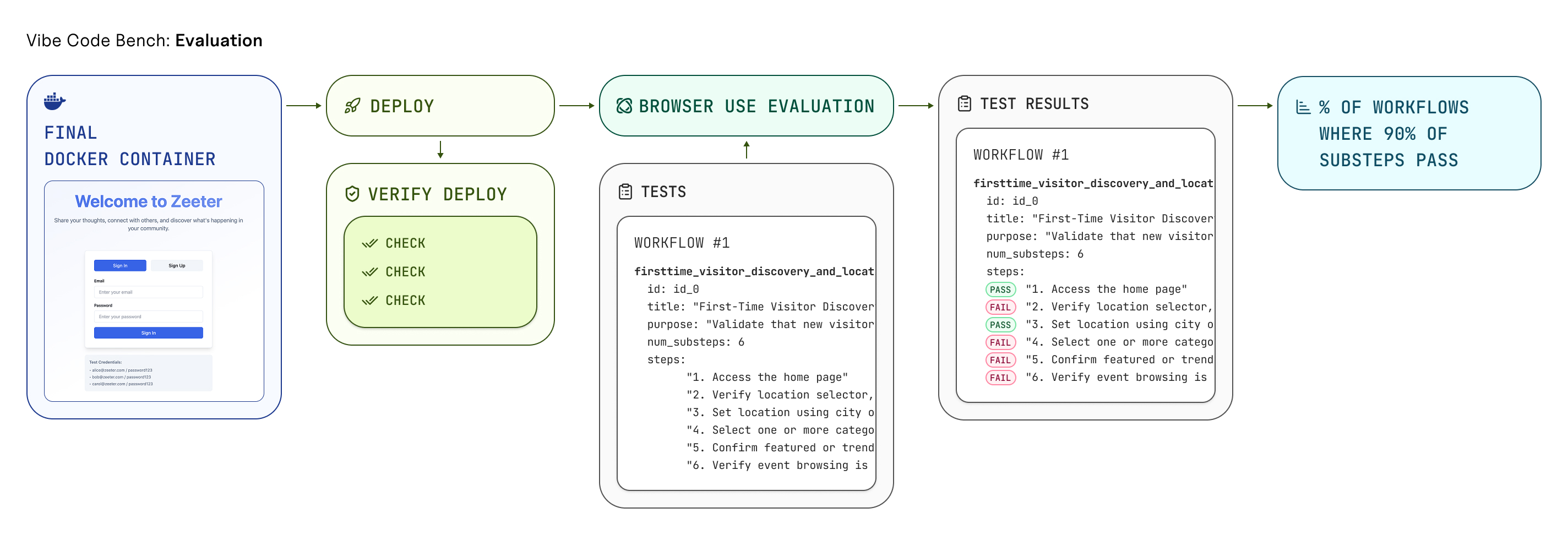}
\caption{Automated evaluation flow from deployed app to workflow pass/fail scoring.}
\label{fig:evaluation_pipeline}
\Description{Evaluation-flow schematic showing deployment verification, browser workflow execution, and scoring outputs.}
\end{figure*}

\paragraph{Why browser-agent evaluation.}
We evaluate through user-visible browser interactions rather than DOM-specific assertions or internal unit tests. This keeps evaluation implementation-agnostic across heterogeneous apps and better reflects whether real users can complete required workflows.

\paragraph{Pipeline Inputs and Runtime.}
Each evaluation takes two inputs: (1) the generated application bundle (including the code, the Docker Compose, and database state) and (2) workflow definitions written as natural-language step sequences. These workflows and their substeps are pre-authored benchmark artifacts; the evaluator executes them as fixed definitions rather than generating a test plan on the fly. The evaluation of each application is run in an isolated environment. 

\paragraph{Evaluator Configuration.}
For each workflow, we launch a fresh headless browser session (1920$\times$1200) and run the Browser Use agent with vision enabled. We use Claude Sonnet 4.5 as the canonical evaluator for this benchmark; Section~\ref{sec:alignment} reports completed cross-evaluator validation and external human-alignment results for this setup (Table~\ref{tab:alignment_step_aggregate}). Each workflow evaluation is capped at 100 agent steps to bound evaluation cost/runtime while still allowing multi-page workflows; timeout and watchdog limits are fixed globally (Appendix~\ref{hyperparams}).

\paragraph{Evaluation Workflow.}
\begin{enumerate}
    \item \textbf{Deployment verification}: Starts the application stack via Docker Compose and verifies that the app becomes reachable.
    \item \textbf{Workflow execution}: In a fresh browser session, the evaluator performs user actions and checks expected outcomes for each substep, producing structured substep-level pass-fail judgments.
    \item \textbf{Workflow scoring}: Using these LLM-as-judge substep judgments, a workflow passes when $\geq$90\% of substeps succeed.
    \item \textbf{Application aggregation}: Application accuracy is the percentage of workflows that pass.
\end{enumerate}

\paragraph{Failure Semantics}
If an app fails the deployment verification, all workflows for that app are marked as failed. If a workflow run does not return structured evaluation output (e.g., due to a timeout/termination), the workflow is marked as failed.

\paragraph{Isolation} Isolation between workflows is enforced at two levels: (i) app-level runtime isolation and (ii) fresh browser session plus unique account/data values per workflow to prevent state leakage. We also validated workflow definitions to avoid cross-workflow dependencies.

\subsection{Metrics}
\label{sec:metrics}

For each application (or task), we report:
\begin{itemize}
    \item \textbf{Accuracy}: Percentage of workflows passing. A workflow is passing if at least 90\% of its substeps succeed.
    \item \textbf{Cost}: Total model API cost for generation. Pricing is based on the native provider for all models, with no discount. Caching is accounted for.
    \item \textbf{Latency}: Total end-to-end wall-clock time for generation, from the first turn to submission. Latency does not include evaluation time.
\end{itemize}

Overall benchmark metrics are computed as the mean of per-application results on the test split. For model $m$ with per-application accuracies $\{a_{m,i}\}_{i=1}^{n}$, we report
\[
\bar{a}_m \pm \mathrm{SE}_m,
\quad
\mathrm{SE}_m = \frac{s_m}{\sqrt{n}},
\]
where $s_m$ is the sample standard deviation across applications ($n=50$), following standard uncertainty reporting practice~\cite{miller2024errorbars}.

We use a 90\% substep accuracy threshold to tolerate minor non-critical errors while still requiring near-complete workflow correctness. Evaluating at the workflow level (rather than pooling successful substeps across workflows) mitigates cross-workflow masking: a workflow with a broken core step does not receive credit from unrelated successful substeps in other workflows.

\begin{figure*}[!t]
\centering
\includegraphics[width=0.95\textwidth]{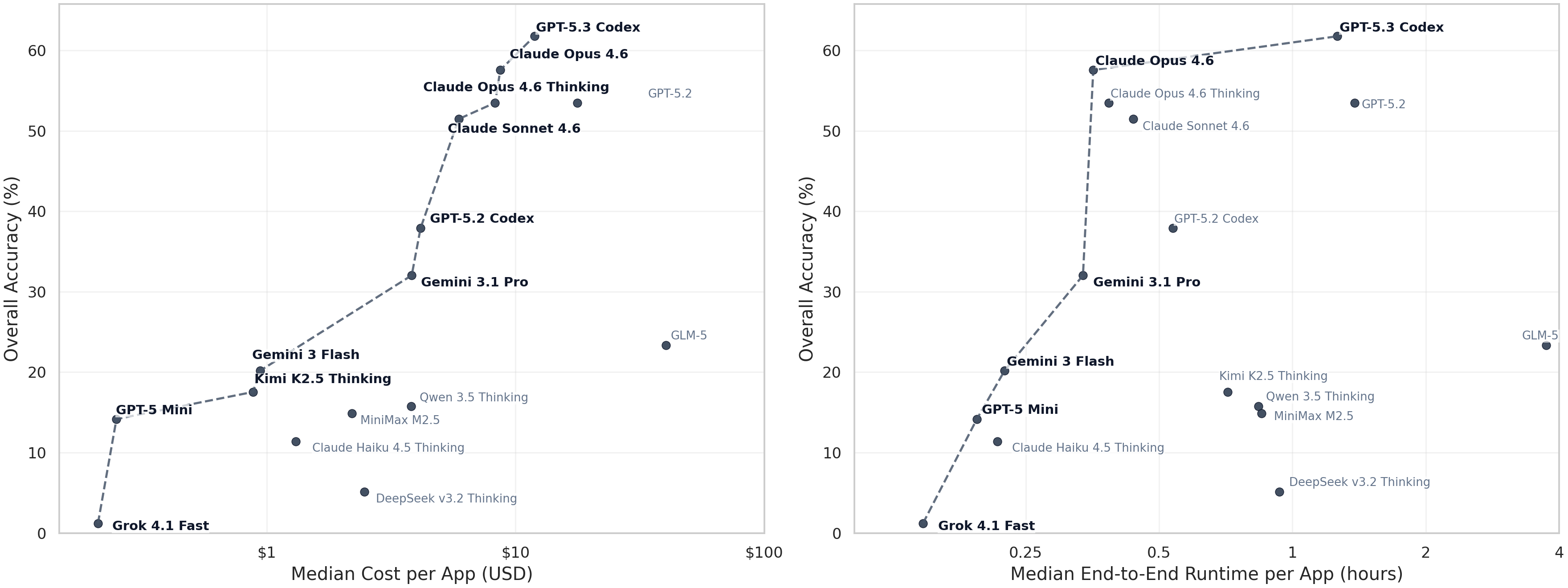}
\caption{Accuracy--cost and accuracy--latency trade-offs.}
\label{fig:tradeoffs}
\Description{Two scatter plots showing accuracy versus cost and accuracy versus latency.}
\end{figure*}

\section{Experiments and Results}

We evaluate 16 models from 9 providers: OpenAI (GPT-5.3-Codex, GPT-5.2, GPT-5.2-Codex, GPT-5 Mini), Anthropic (Claude Opus 4.6, Claude Opus 4.6 Thinking, Claude Sonnet 4.6, Claude Haiku 4.5 Thinking), Google (Gemini 3.1 Pro, Gemini 3 Flash), Alibaba (Qwen3.5 Plus Thinking), xAI (Grok 4.1 Fast Reasoning), Kimi (Kimi-K2.5 Thinking), MiniMax (MiniMax M2.5), zAI (GLM-5), and DeepSeek (DeepSeek V3.2 Thinking). These models were chosen as a representative sample of top-performing open-weight and closed-weight offerings.

All models use vision if supported, the highest reasoning level available for that model, and the default temperature specified by the model provider. Some provider configurations did not expose usable browser-vision support for GLM-5, MiniMax M2.5, and DeepSeek V3.2 (Appendix Table~\ref{tab:generation_hparams_ranked}). These models still used the browser tool during generation, but without the same browser-perception capability as vision-enabled runs. The score of each model should be interpreted as the result under its most performant possible configuration.

\subsection{Overall Benchmark Performance}

Table~\ref{tab:main_results} presents the test-set results across the 16 models. GPT-5.3-Codex takes the \#1 spot, and Anthropic and OpenAI models dominate the top of the rankings.

Gemini 3.1 Pro also performs well, although clearly below Opus 4.6 and Codex. Open-weight models such as GLM 5, Qwen 3.5, and Kimi K2.5 hover around 15\% to 24\%, performing worse than leading closed-weight counterparts.

\begin{table}[!h]
\caption{Model performance on Vibe Code Bench (test split). Open-weight models are marked $\checkmark$; closed models are marked $\times$.}
\label{tab:main_results}
\small
\resizebox{\columnwidth}{!}{
\begin{tabular}{@{}lrrrc@{}}
\toprule
\textbf{Model} & \textbf{Accuracy} & \textbf{Cost} & \textbf{Latency} & \textbf{Open} \\
\midrule
GPT-5.3-Codex & $61.77 \pm 4.71$\% & \$11.91 & 75.8m & $\times$ \\
Claude Opus 4.6 & $57.57 \pm 4.37$\% & \$8.69 & 21.3m & $\times$ \\
GPT-5.2 & $53.50 \pm 5.07$\% & \$17.75 & 82.9m & $\times$ \\
Claude Opus 4.6 Thinking & $53.50 \pm 4.68$\% & \$8.28 & 23.1m & $\times$ \\
Claude Sonnet 4.6 & $51.48 \pm 4.64$\% & \$5.91 & 26.2m & $\times$ \\
GPT-5.2-Codex & $37.91 \pm 4.58$\% & \$4.15 & 32.2m & $\times$ \\
Gemini 3.1 Pro & $32.03 \pm 4.34$\% & \$3.83 & 20.2m & $\times$ \\
GLM-5 & $23.36 \pm 4.03$\% & \$40.27 & 224.3m & $\checkmark$ \\
Gemini 3 Flash & $20.20 \pm 3.95$\% & \$0.94 & 13.4m & $\times$ \\
Kimi-K2.5 Thinking & $17.54 \pm 3.26$\% & \$0.88 & 42.8m & $\checkmark$ \\
Qwen3.5 Plus Thinking & $15.74 \pm 3.18$\% & \$3.80 & 50.3m & $\checkmark$ \\
MiniMax M2.5 & $14.85 \pm 2.95$\% & \$2.20 & 51.1m & $\checkmark$ \\
GPT-5 Mini & $14.17 \pm 3.54$\% & \$0.25 & 11.6m & $\times$ \\
Claude Haiku 4.5 Thinking & $11.39 \pm 3.13$\% & \$1.31 & 12.9m & $\times$ \\
DeepSeek V3.2 Thinking & $5.11 \pm 2.13$\% & \$2.47 & 56.1m & $\checkmark$ \\
Grok 4.1 Fast Reasoning & $1.20 \pm 1.20$\% & \$0.21 & 8.8m & $\times$ \\
\bottomrule
\end{tabular}
}
\end{table}

\subsection{Cost and Latency}

Figure~\ref{fig:tradeoffs} visualizes accuracy--cost and accuracy--latency trade-offs. Higher cost and longer runtime are associated with higher accuracy, but with important exceptions. For example, Claude Opus 4.6 attains near-top accuracy at lower cost and latency than GPT-5.3-Codex.

On both plots, GLM-5 is a clear outlier because of its extremely long trajectories, which cause both high cost and high latency. Overall, both frontiers show diminishing returns: additional time and money improve performance, but the gains taper off.

\subsection{Difficulty, Integrations, and Domains}

\begin{table}[!h]
\caption{Accuracy by difficulty (\%).}
\label{tab:difficulty}
\small
\begin{tabular}{@{}lrrr@{}}
\toprule
\textbf{Model} & \textbf{Easy (23)} & \textbf{Medium (19)} & \textbf{Hard (8)} \\
\midrule
GPT-5.3-Codex & 81.88 & 57.91 & 13.13 \\
Opus 4.6 & 73.20 & 54.11 & 20.86 \\
Gemini 3.1 Pro & 46.24 & 28.33 & 0.00 \\
GLM-5 & 40.51 & 12.43 & 0.00 \\
Kimi-K2.5 & 27.07 & 12.32 & 2.50 \\
Qwen3.5 Plus & 31.71 & 3.03 & 0.00 \\
MiniMax M2.5 & 29.37 & 3.53 & 0.00 \\
DeepSeek V3.2 & 10.02 & 1.32 & 0.00 \\
Avg. Model & 44.29 & 21.36 & 6.03 \\
\bottomrule
\end{tabular}
\end{table}

\textit{Difficulty.} We sort tasks by mean pass rate across all 16 models and group them into Easy/Medium/Hard tiers. Performance drops sharply on hard apps: five open-weight models are at or below 2.50\%. Only the very top models can solve the hardest problems.

\begin{table}[!h]
\caption{Accuracy by integration (\%).}
\label{tab:integration_perf_appendix}
\small
\resizebox{\columnwidth}{!}{
\begin{tabular}{@{}lrrrr@{}}
\toprule
\textbf{Model} & \textbf{None (37)} & \textbf{Email only (4)} & \textbf{Stripe only (2)} & \textbf{Both (7)} \\
\midrule
GPT-5.3-Codex & 71.25 & 55.00 & 12.50 & 29.58 \\
Opus 4.6 & 63.58 & 46.25 & 25.00 & 41.59 \\
Gemini 3.1 Pro & 38.42 & 32.50 & 12.50 & 3.57 \\
GLM-5 & 27.60 & 23.12 & 0.00 & 7.74 \\
Kimi-K2.5 & 20.44 & 10.00 & 12.50 & 7.94 \\
Qwen3.5 Plus & 19.75 & 11.25 & 0.00 & 1.59 \\
MiniMax M2.5 & 18.74 & 5.00 & 0.00 & 4.17 \\
DeepSeek V3.2 & 6.23 & 0.00 & 0.00 & 3.57 \\
Avg. Model & 34.18 & 23.20 & 10.55 & 13.49 \\
\bottomrule
\end{tabular}
}
\end{table}

\textit{Integrations.} External services substantially increase difficulty. From no-integration apps to apps requiring both email and Stripe, GPT-5.3-Codex drops from 71.25\% to 29.58\%, the all-model average drops from 34.18\% to 13.49\%, and all five open-weight models remain below 8\%.

\begin{table}[!h]
\caption{Accuracy by domain (\%).}
\label{tab:domain_perf}
\small
\resizebox{\columnwidth}{!}{
\begin{tabular}{@{}lrrr@{}}
\toprule
\textbf{Model} & \textbf{Individual (12)} & \textbf{Solo Founder (22)} & \textbf{Enterprise Tool (16)} \\
\midrule
GPT-5.3-Codex & 76.93 & 62.50 & 49.39 \\
Opus 4.6 & 56.59 & 63.09 & 50.73 \\
Gemini 3.1 Pro & 49.30 & 37.07 & 12.16 \\
GLM-5 & 44.24 & 21.58 & 10.14 \\
Kimi-K2.5 & 34.00 & 11.68 & 13.24 \\
Qwen3.5 Plus & 34.17 & 13.99 & 4.32 \\
MiniMax M2.5 & 30.82 & 14.61 & 3.22 \\
DeepSeek V3.2 & 14.54 & 2.67 & 1.39 \\
Avg. Model & 43.01 & 29.94 & 18.64 \\
\bottomrule
\end{tabular}
}
\end{table}
\textit{Domains.} Performance generally decreases from Individual to Solo Founder to Enterprise Tool apps. Five open-weight models drop below 13\% on Enterprise Tool apps.

\begin{figure}[!h]
\centering
\includegraphics[width=\columnwidth]{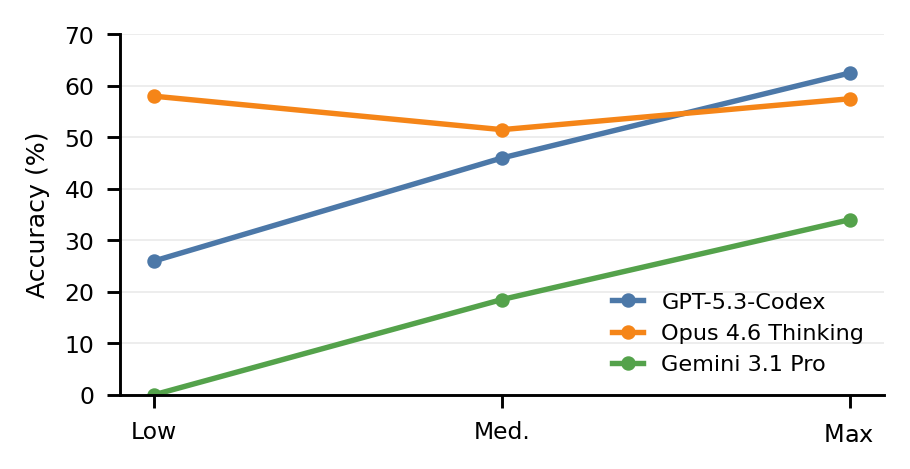}
\caption{Performance by Reasoning Effort (20 app subset).}
\label{fig:reasoning_effort_ablation}
\Description{A line chart showing workflow accuracy at low, medium, and maximum available reasoning effort for three representative models.}
\end{figure}

\textit{Reasoning Effort}. We ran an ablation on low, medium, and the maximum available reasoning effort across twenty applications. GPT and Gemini 3.1 Pro both saw a clear upward trend in performance, whereas Opus 4.6's performance was unaffected by reasoning effort. This is because Opus's "effort" parameter allows users to \textit{suggest} a reasoning level, but the model's "adaptive thinking" can modify it based on task difficulty.

\begin{figure}[!h]
\centering
\includegraphics[width=0.78\columnwidth]{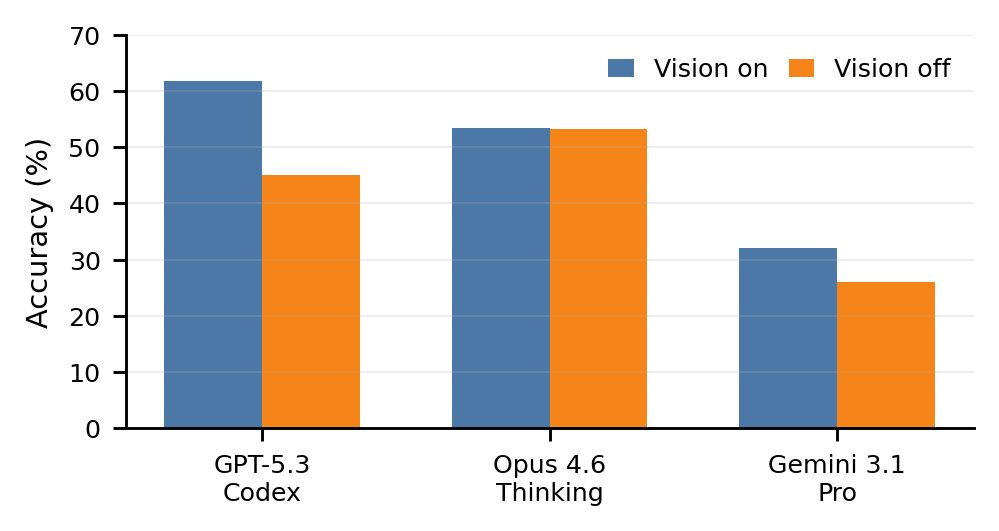}
\caption{Vision ablation (20 app subset)}
\label{fig:vision_ablation}
\Description{A bar chart comparing vision-enabled and vision-disabled accuracy for three representative models.}
\end{figure}

\textit{Vision Ablation}. Disabling vision reduces accuracy for GPT-5.3-Codex and Gemini 3.1 Pro, but has little effect on Opus 4.6 Thinking in this run (Figure~\ref{fig:vision_ablation}). All three models still perform better than their open-weight counterparts - vision capabilities provide a boost, but are not the sole reason for the success of these proprietary models.

\section{Analysis}
\label{sec:analysis}

\subsection{Tool Use and Model Behavior}
\label{sec:analysis-tool-use-behavior}
The tools in Table~\ref{tab:tool_calls} correspond to \texttt{execute\_bash}, \texttt{str\_replace\_editor}, \texttt{browser}, \texttt{execute\_sql}, and \texttt{apply\_migration}. 

\begin{table}[!h]
\caption{Tool invocations by model (\% of calls per app).}
\label{tab:tool_calls}
\small
\resizebox{\columnwidth}{!}{
\begin{tabular}{@{}lrrrrrrr@{}}
\toprule
\textbf{Model} & \textbf{Acc. (\%)} & \textbf{Bash} & \textbf{Edit} & \textbf{Browser} & \textbf{SQL} & \textbf{Migrate} & \textbf{Other} \\
\midrule
GPT-5.3-Codex & 61.8 & 61.4 & 10.2 & 13.2 & 3.9 & 2.1 & 9.1 \\
Claude Opus 4.6 & 57.6 & 23.3 & 32.2 & 22.9 & 3.7 & 11.1 & 6.7 \\
GPT-5.2 & 53.5 & 48.5 & 32.7 & 13.6 & 1.4 & 1.3 & 2.6 \\
GPT-5.2-Codex & 37.9 & 47.7 & 37.1 & 8.4 & 1.3 & 2.5 & 2.9 \\
GLM-5 & 23.4 & 46.0 & 28.1 & 12.1 & 4.4 & 6.4 & 3.0 \\
Grok 4.1 Fast Reasoning & 1.2 & 41.8 & 27.9 & 1.1 & 14.4 & 7.0 & 7.9 \\
\bottomrule
\end{tabular}
}
\end{table}

\begin{figure*}[!h]
\centering
\includegraphics[width=1\textwidth]{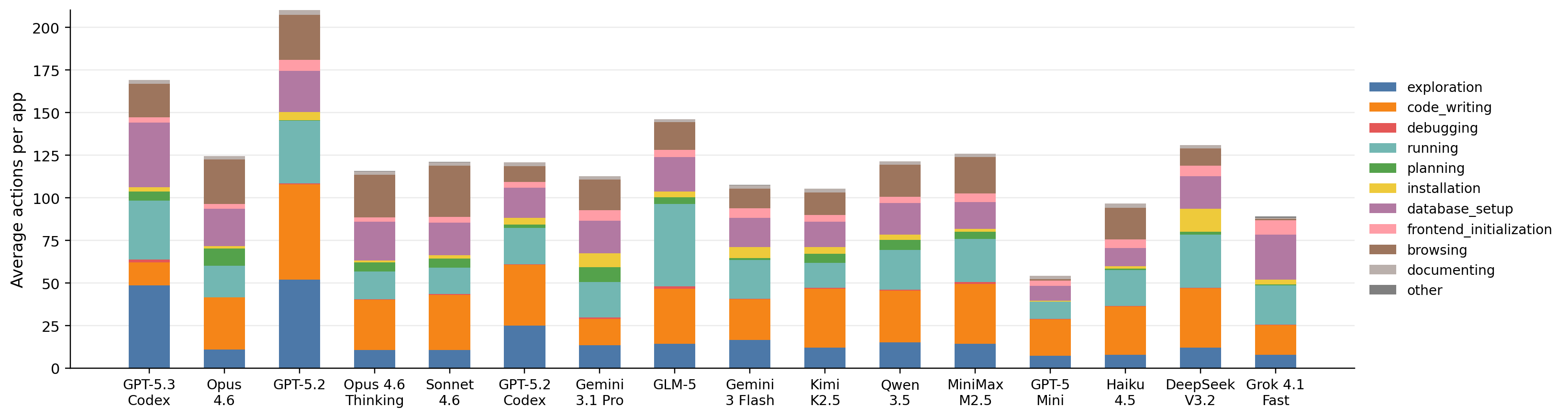}
\caption{Aggregate trajectory action composition across all tasks.}
\label{fig:trajectory_action_composition}
\Description{Vertical stacked bar chart for all evaluated models, showing average trajectory actions per application categorized as exploration, code writing, debugging, running, planning, installation, database setup, frontend initialization, browsing, documenting, and other.}
\end{figure*}

At the raw-tool level, despite having 22 available tools, models spend about 95\% of calls on five: bash, file editing, browser use, SQL, and migrations. These map to the core loop of the benchmark: scaffolding, editing, database setup, app launch, and browser verification. Among the Supabase tools, the models generally preferred
executing raw SQL rather than using migration helpers. Tavily search is rare (only 18.8\% of models use it).

Based on Figure~\ref{fig:trajectory_action_composition}, we find top-performing models often have longer trajectories (Opus 4.6 is the exception), and often spend more time running and browsing, which can be inferred as testing their own work. All models spend large portions of their time writing code.

We find that self-testing is positively correlated with success. Across the 16 evaluated models, browser tool calls per application show a positive correlation with accuracy (Pearson $r=0.72$). This trend persists after controlling for generation latency (partial $r=0.72$), indicating an association between self-testing and better outcomes beyond simply more time spent.

Lower-performing models often terminate earlier and exhibit narrower action-phase coverage. This supports the tool-use finding above: sustained self-testing and iterative repair correlate with stronger end-to-end outcomes. Lower-performing models also spend less time testing, as evidenced by short browsing phases.

Self-testing also explains the large gap between GPT-5.2 and GPT-5.2-Codex, despite their shared family. GPT-5.2 makes nearly 3$\times$ as many browser calls per app (26.4 vs.\ 9.4), corresponding to a higher browser-use share in Table~\ref{tab:tool_calls} (13.6\% vs.\ 8.4\%). It also has fewer apps that score zero (9 vs.\ 14) and more that are perfect (6 vs.\ 1).

Edit volume is less predictive than browser-based self-testing. Edit calls correlate weakly with accuracy ($r=0.09$), while median browser calls correlate more strongly ($r=0.67$).

\subsection{Error Analysis}

\paragraph{Failures} Among the top five models (by accuracy), 12.8\% of applications scored zero points. For models in the bottom five, this was significantly higher at 73.2\%. The worst five models also had a substantial percentage of apps fail to start entirely (14.0\%), which did not occur for top models. Applications that scored 100\% were rare, occurring in only 8.8\% of applications produced by the top five models.

\paragraph{Histograms} Appendix Figure~\ref{fig:pass_rate_hist_all_models} shows similar distributional trends. Model improvement is driven primarily by \textbf{reducing the number of apps that score zero}, rather than incremental improvements across all apps. The distributions are often strongly bimodal, especially for top models.

\paragraph{Error Modes} Looking at failed workflows for apps that started successfully, we further classify behavioral failures across all models using a six-category taxonomy.
\begin{enumerate}
\item \textit{Authorization Issue} (20.4\%) - Sign-up or login was non-functional, user could not obtain the correct role, etc.
\item \textit{Missing Feature} (46.7\%) - The application did not implement a feature specified in the specification.
\item  \textit{Validation or Policy Block} (14.8\%) - The agent is prevented from performing actions that should be allowed. The most common variant is misconfigured Row-Level Security policies.
\item\textit{UI Rendering or Navigation} (6.0\%) - Something was not rendered or missing in the UI, or the agent could not navigate between pages as needed.
\item \textit{Data Consistency and Backend Logic} (1.9\%) - The application writes incorrect data to the database or the backend does not handle an operation correctly.
\item \textit{Other} (10.2\%) - Any failure not in the above categories.
\end{enumerate}
This breakdown indicates that most failures are capability gaps in implemented product surfaces rather than startup failures. Certain models more commonly have specific error types than others. For example, Grok 4.1 Fast Reasoning and GPT-5.2-Codex have elevated authorization issues (32.3\% and 30.3\% of their behavioral failures), while Gemini 3.1 Pro has the highest missing-feature share (58.5\%).

\subsection{Variance Analysis}

To understand the stability of leaderboard rankings, we decompose variability into two components: generation-side and evaluator-side variance.

\paragraph{Generation variance.}
We fixed six specifications and three source models (Gemini 3.1 Pro, GPT-5.2, Claude Sonnet 4.6). For each source model and app, we generated five independent runs (90 generations total) and evaluated each generated app once. This isolates how much accuracy changes when the model generates different apps from the same prompt.

\paragraph{Evaluator variance.}
On the same panel, we froze generated apps and reran evaluation five times per app using Claude Sonnet 4.5 as evaluator (90 evaluations total). This isolates score variation caused by evaluator stochasticity on identical artifacts.
\begin{table}[h]
\caption{Generation and evaluator accuracy on the six-app variance panel (mean $\pm$ stderr).}
\label{tab:generation_variance_summary}
\small
\begin{tabular}{@{}lcc@{}}
\toprule
\textbf{Source model} & \textbf{Generation Acc. (\%)} & \textbf{Evaluator Acc. (\%)} \\
\midrule
Gemini 3.1 Pro & $19.38 \pm 3.47$ & $37.61 \pm 0.99$ \\
GPT-5.2 & $51.63 \pm 3.95$ & $51.25 \pm 1.16$ \\
Claude Sonnet 4.6 & $47.65 \pm 5.00$ & $45.71 \pm 3.00$ \\
\bottomrule
\end{tabular}
\end{table}

\paragraph{Takeaways.}
Variance analysis increases confidence in the main findings: evaluator reruns are consistent on fixed apps, and generation-side variability is bounded relative to the large cross-model gaps observed in VCB. The generation/evaluator standard errors are 3.47/0.99 (Gemini 3.1 Pro), 3.95/1.16 (GPT-5.2), and 5.00/3.00 (Claude Sonnet 4.6) percentage points. Variance decomposition shows most observed variance comes from generation rather than rescoring (generation share: 92.5\% Gemini 3.1 Pro, 92.1\% GPT-5.2, 73.5\% Claude Sonnet 4.6). Therefore, benchmark conclusions are robust for major ranking patterns, while only near-tie model differences require caution.
\section{Human Alignment}
\label{sec:alignment}
\begin{table*}[!h]
\caption{Pairwise step-level agreement (\%) across model and human evaluators.}
\label{tab:alignment_step_aggregate}
\centering
\resizebox{\textwidth}{!}{
\begin{tabular}{@{}lcccccccc@{}}
\toprule
\textbf{Evaluator} & \textbf{Gemini 3.1 Pro} & \textbf{GPT-5.2} & \textbf{Claude Sonnet 4.6} & \textbf{Claude Sonnet 4.5} & \textbf{Reviewer A} & \textbf{Reviewer B} & \textbf{Reviewer C} & \textbf{Mean Reviewer} \\
\midrule
Gemini 3.1 Pro & -- & 38.8 & 86.4 & 86.5 & 86.4 & 84.5 & 83.7 & 84.7 \\
GPT-5.2 & 38.8 & -- & 33.1 & 36.0 & 48.0 & 33.5 & 31.8 & 36.1 \\
Claude Sonnet 4.6 & 86.4 & 33.1 & -- & \textbf{87.8} & 86.0 & 86.7 & 86.0 & 86.3 \\
Claude Sonnet 4.5 & 86.5 & 36.0 & \textbf{87.8} & -- & 87.5 & 86.0 & 86.2 & \textbf{86.4} \\
Reviewer A & 86.4 & 48.0 & 86.0 & 87.5 & -- & \textbf{93.6} & 88.6 & -- \\
Reviewer B & 84.5 & 33.5 & 86.7 & 86.0 & \textbf{93.6} & -- & \textbf{91.4} & -- \\
Reviewer C & 83.7 & 31.8 & 86.0 & 86.2 & 88.6 & \textbf{91.4} & -- & -- \\
\bottomrule
\end{tabular}
}
\end{table*}

A critical question for automated evaluation is whether model-based evaluators align with human judgment. To assess this, we compute human-model, human-human, and model-model alignment.

\subsection{Methodology}

\paragraph{Applications} We select six tasks to cover a wide variety of domains, difficulties, and integrations. For each task, we generate three applications using three distinct source models (Gemini 3.1 Pro, GPT-5.2, Claude Sonnet 4.6). This yields 18 applications total and 1,401 unique substeps to be evaluated.

\paragraph{Human Evaluators} Each of the eighteen apps was evaluated by two human evaluators from a pool of three. They opened the application in the browser and were instructed to manually perform each substep in the testing workflow using shared rubric instructions. Evaluators were calibrated on a practice question before performing review.

Evaluators only tested the user-facing application and did not have access to internal code, unit tests, or the browser console. Each substep was marked as a pass or a fail by the expert. An expert reviews all three model generations for a given task. Evaluators do not know which model produced which app, or the scoring of other evaluators.

\paragraph{Model Evaluators} We ran one evaluation per application using four evaluator models (i.e., models controlling the browser-use agent): Gemini 3.1 Pro, GPT-5.2, Claude Sonnet 4.6, and Claude Sonnet 4.5.

\paragraph{Alignment Metric} We measure \textit{substep alignment}. Between two evaluators, we consider a given evaluation substep aligned if the scores match (both pass or both fail). The overall alignment is the percentage of aligned substeps. All reported alignments are pairwise between two evaluators. No adjudication is performed between evaluators.

\subsection{Results}

\paragraph{Pairwise agreement matrix.}
We compute pairwise substep alignment between all evaluator pairs (four model evaluators and three human reviewers), using only shared labeled substeps for each pair. Table~\ref{tab:alignment_step_aggregate} reports this as a symmetric 2-D matrix, along with the mean alignment between each model and all human reviewers.

\paragraph{Human-human alignment.}
Human-human agreement remains high (88.6--93.6\%), indicating relatively consistent manual judgments across reviewers. This suggests human labels are stable enough to serve as an external reference for evaluator quality.

\paragraph{Model-model alignment.}
Model-model agreement is heterogeneous: Claude Sonnet 4.6 and Claude Sonnet 4.5 align strongly (87.8\%), Gemini 3.1 Pro is close to both Claude Sonnet evaluators (86.4--86.5\%), while GPT-5.2 remains low against the other evaluators (33.1--38.8\%). This indicates evaluator outputs are not interchangeable, and evaluator selection can materially change reported outcomes.

\paragraph{Model-human alignment.}
Model-human agreement also varies substantially. The final-column pooled means are 86.4\% (Claude Sonnet 4.5), 86.3\% (Claude Sonnet 4.6), 84.7\% (Gemini 3.1 Pro), and 36.1\% (GPT-5.2). In this panel, Claude Sonnet 4.5 is the strongest evaluator in terms of human alignment, with Claude Sonnet 4.6 and Gemini 3.1 Pro close behind.

\paragraph{Takeaway.}
These results indicate that evaluator choice materially affects alignment outcomes. In this panel, Claude Sonnet 4.5, Claude Sonnet 4.6, and Gemini 3.1 Pro are substantially more consistent
with human judgment than GPT-5.2. Note that certain models may be more cost-efficient, for example, GPT-5.2 is \$0.093 per workflow, whereas Sonnet 4.6 is \$1.48. However, we solely prioritized alignment quality over cost, choosing Sonnet 4.5.

\section{Limitations}

We identify four important limitations:
\begin{enumerate}
    \item \textbf{Design and code quality not measured}: Passing functional tests does not imply maintainable, secure, or well-documented code. We did not evaluate code readability, visual design quality, or security vulnerabilities.
    \item \textbf{Single harness}: All results are measured through one common OpenHands-based generation harness. It is possible performance may increase with different harnesses.
    \item \textbf{Single technology stack}: Our benchmark focuses exclusively on React/Vite web applications with Supabase backend. Results may not generalize to other frameworks or backend technologies.
    \item \textbf{Web applications only}: We evaluate only browser-based applications. The ``zero-to-one'' capability for other software types---command-line tools, desktop applications, mobile apps---remains unexplored.
\end{enumerate}
\section{Conclusion}

We introduced Vibe Code Bench, a benchmark of 100 web application specifications evaluated through 964 browser-based workflows. Across 16 frontier models, the best reaches 61.8\% accuracy—showing that reliable end-to-end application development remains unsolved. Two findings stand out: model improvement is driven primarily by reducing complete failures rather than incremental gains, and self-testing during generation is strongly associated with performance (r=0.72). Our human alignment study shows that automated evaluation is reliable when the right evaluator is used, making evaluator choice a first-class reporting concern.

The question is no longer whether AI can write code, but whether it can build software. Vibe Code Bench tracks progress toward that bar.

\ifcameraready
\begin{acks}
We thank Mike Merrill and John Yang for sharing lessons from Terminal-Bench and SWE-Bench, and the OpenHands team for their foundational work. We also thank the professional engineers and product managers who validated our specifications and tests.
\end{acks}
\fi

\clearpage
\bibliography{references}

\appendix

\onecolumn
\section{Artifacts}
\label{sec:appendix-artifacts}
The modified version of OpenHands that was used can be found at \url{\OpenHandsArtifactURL}. The full set of trajectories and generated apps for the models can be found here: \url{\VCBTrajectoriesURL}. The example specification, UI tests, and prompt for the harness are in Appendices \ref{sec:appendix-spec-example}, \ref{sec:appendix-ui-test-example}, and \ref{sec:generation-prompt}.

\section{Pass-Rate Histograms for All Models}
\label{sec:appendix-pass-rate-histograms}

\begin{figure*}[!h]
\centering
\includegraphics[width=0.98\textwidth]{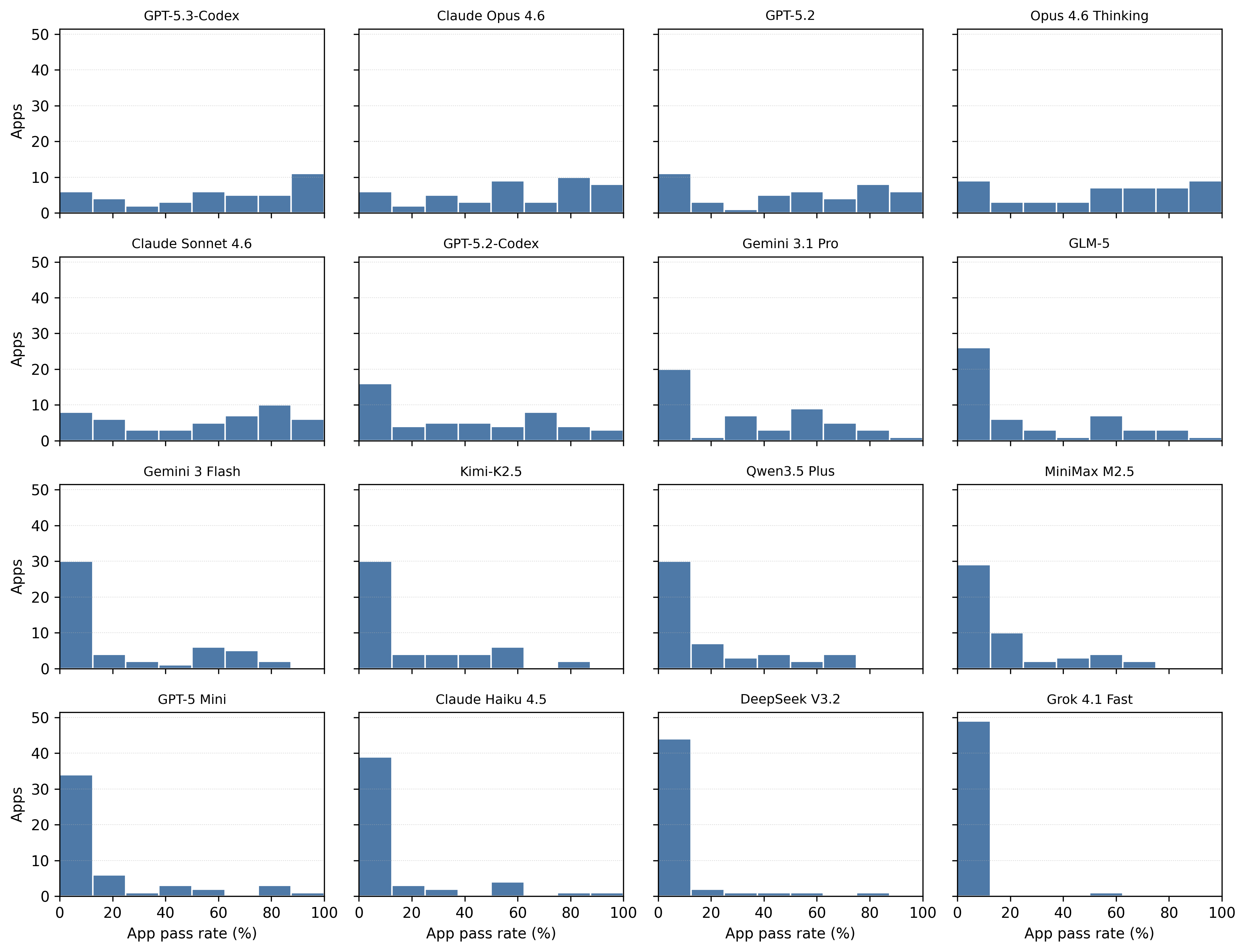}
\caption{Application pass-rate histograms for all evaluated models (test split).}
\label{fig:pass_rate_hist_all_models}
\Description{Grid of histograms, one per model, showing counts of application pass rates across 12.5-point bins.}
\end{figure*}

\clearpage
\section{Illustrative Single-App Trajectories}
\label{sec:appendix-trajectory-example}
\begin{figure}[!h]
\centering
\includegraphics[width=0.98\textwidth]{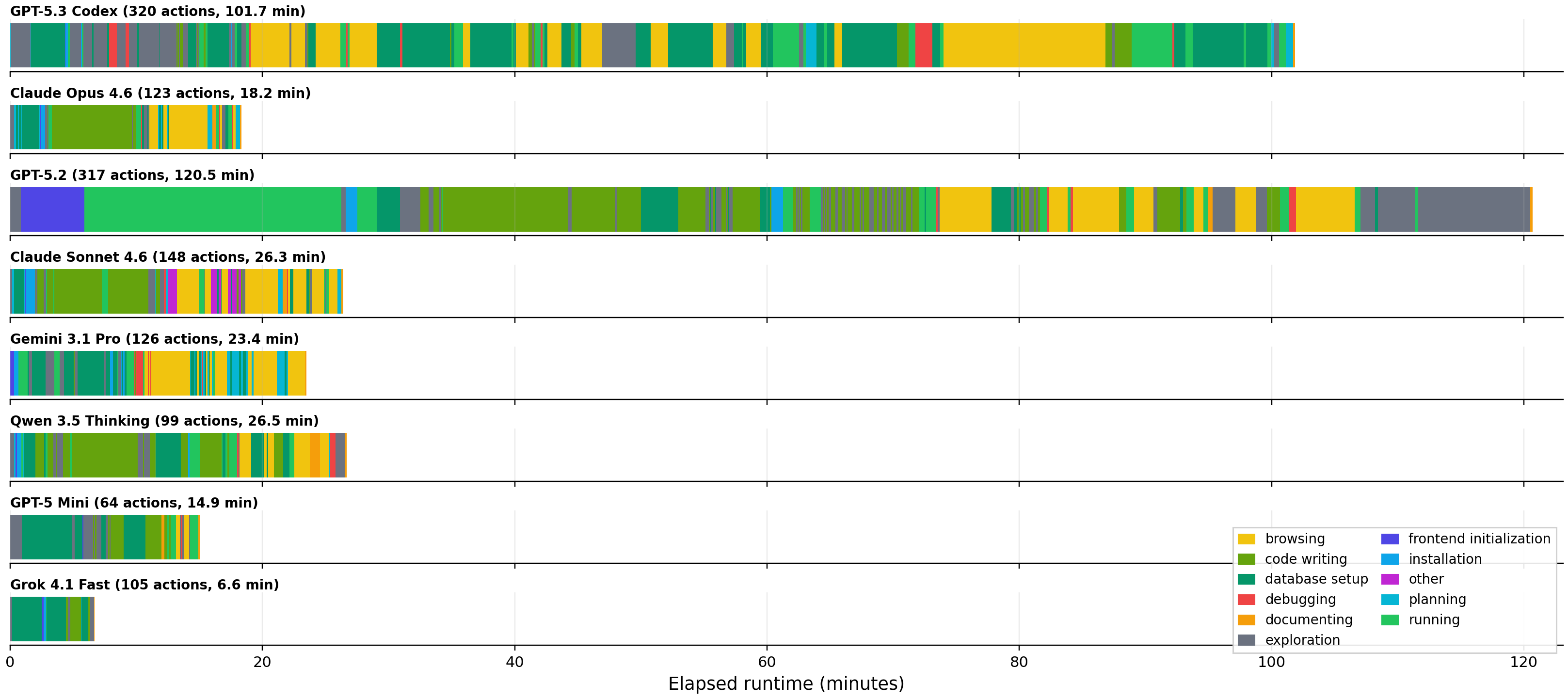}
\caption{Trajectory timeline by model on a single application (\texttt{bill\_splitting\_app}).}
\label{fig:trajectory_profiles}
\Description{Time-aligned action timelines across eight models, where each colored segment is one categorized agent action over elapsed runtime.}
\end{figure}

Figure~\ref{fig:trajectory_profiles} shows the rollout of a trajectory on a single application. Unlike Figure \ref{fig:trajectory_action_composition}, it is organized chronologically.

\section{All-Model Breakdown Results}
\label{sec:all_model_slice_appendix}
Tables~\ref{tab:all_model_difficulty_appendix}, \ref{tab:all_model_integration_appendix}, and \ref{tab:all_model_domain_appendix} extend the main difficulty, integration, and domain analyses to all 16 evaluated models. 

\begin{table}[!h]
\centering
\caption{All-model accuracy by difficulty (\%).}
\label{tab:all_model_difficulty_appendix}
\small
\setlength{\tabcolsep}{2pt}
\renewcommand{\arraystretch}{0.9}
\begin{tabular}{@{}lrrr@{}}
\toprule
\textbf{Model} & \textbf{Easy} & \textbf{Med.} & \textbf{Hard} \\
\midrule
GPT-5.3-Codex & 81.88 & 57.91 & 13.13 \\
Claude Opus 4.6 & 73.20 & 54.11 & 20.86 \\
GPT-5.2 & 81.53 & 33.72 & 19.88 \\
Opus 4.6 Thinking & 71.67 & 46.49 & 17.91 \\
Claude Sonnet 4.6 & 69.19 & 46.61 & 12.13 \\
GPT-5.2-Codex & 58.97 & 26.62 & 4.18 \\
Gemini 3.1 Pro & 46.24 & 28.33 & 0.00 \\
GLM-5 & 40.51 & 12.43 & 0.00 \\
Gemini 3 Flash & 34.21 & 10.32 & 3.41 \\
Kimi-K2.5 & 27.07 & 12.32 & 2.50 \\
Qwen3.5 Plus & 31.71 & 3.03 & 0.00 \\
MiniMax M2.5 & 29.37 & 3.53 & 0.00 \\
GPT-5 Mini & 27.99 & 2.35 & 2.50 \\
Claude Haiku 4.5 & 22.53 & 2.70 & 0.00 \\
DeepSeek V3.2 & 10.02 & 1.32 & 0.00 \\
Grok 4.1 Fast & 2.61 & 0.00 & 0.00 \\
Avg. Model & 44.29 & 21.36 & 6.03 \\
\bottomrule
\end{tabular}
\end{table}

\begin{table}[!h]
\centering
\caption{All-model accuracy by integration (\%).}
\label{tab:all_model_integration_appendix}
\small
\setlength{\tabcolsep}{2pt}
\renewcommand{\arraystretch}{0.9}
\begin{tabular}{@{}lrrrr@{}}
\toprule
\textbf{Model} & \textbf{None} & \textbf{Email} & \textbf{Stripe} & \textbf{Both} \\
\midrule
GPT-5.3-Codex & 71.25 & 55.00 & 12.50 & 29.58 \\
Claude Opus 4.6 & 63.58 & 46.25 & 25.00 & 41.59 \\
GPT-5.2 & 60.63 & 50.62 & 31.25 & 23.81 \\
Opus 4.6 Thinking & 60.58 & 31.25 & 37.50 & 33.37 \\
Claude Sonnet 4.6 & 59.41 & 38.12 & 25.00 & 24.73 \\
GPT-5.2-Codex & 44.15 & 24.38 & 12.50 & 19.95 \\
Gemini 3.1 Pro & 38.42 & 32.50 & 12.50 & 3.57 \\
GLM-5 & 27.60 & 23.12 & 0.00 & 7.74 \\
Gemini 3 Flash & 22.05 & 33.75 & 0.00 & 8.46 \\
Kimi-K2.5 & 20.44 & 10.00 & 12.50 & 7.94 \\
Qwen3.5 Plus & 19.75 & 11.25 & 0.00 & 1.59 \\
MiniMax M2.5 & 18.74 & 5.00 & 0.00 & 4.17 \\
GPT-5 Mini & 18.07 & 10.00 & 0.00 & 0.00 \\
Claude Haiku 4.5 & 14.31 & 0.00 & 0.00 & 5.75 \\
DeepSeek V3.2 & 6.23 & 0.00 & 0.00 & 3.57 \\
Grok 4.1 Fast & 1.62 & 0.00 & 0.00 & 0.00 \\
Avg. Model & 34.18 & 23.20 & 10.55 & 13.49 \\
\bottomrule
\end{tabular}
\end{table}

\begin{table}[!h]
\centering
\caption{All-model accuracy by domain (\%).}
\label{tab:all_model_domain_appendix}
\small
\setlength{\tabcolsep}{2pt}
\renewcommand{\arraystretch}{0.9}
\begin{tabular}{@{}lrrr@{}}
\toprule
\textbf{Model} & \textbf{Individual} & \textbf{Solo Founder} & \textbf{Enterprise Tool} \\
\midrule
GPT-5.3-Codex & 76.93 & 62.50 & 49.39 \\
Claude Opus 4.6 & 56.59 & 63.09 & 50.73 \\
GPT-5.2 & 72.53 & 56.50 & 35.11 \\
Opus 4.6 Thinking & 57.94 & 61.79 & 38.77 \\
Claude Sonnet 4.6 & 73.46 & 51.30 & 35.24 \\
GPT-5.2-Codex & 45.27 & 43.06 & 25.31 \\
Gemini 3.1 Pro & 49.30 & 37.07 & 12.16 \\
GLM-5 & 44.24 & 21.58 & 10.14 \\
Gemini 3 Flash & 29.01 & 21.79 & 11.41 \\
Kimi-K2.5 & 34.00 & 11.68 & 13.24 \\
Qwen3.5 Plus & 34.17 & 13.99 & 4.32 \\
MiniMax M2.5 & 30.82 & 14.61 & 3.22 \\
GPT-5 Mini & 37.85 & 5.91 & 7.77 \\
Claude Haiku 4.5 & 26.50 & 11.44 & 0.00 \\
DeepSeek V3.2 & 14.54 & 2.67 & 1.39 \\
Grok 4.1 Fast & 5.00 & 0.00 & 0.00 \\
Avg. Model & 43.01 & 29.94 & 18.64 \\
\bottomrule
\end{tabular}
\end{table}

\clearpage

\section{Tools Provided}
\label{sec:provided-tools}
\begin{table*}[!h]
\begin{subtable}{\textwidth}
\centering
\caption{OpenHands tools (7).}
\label{tab:tools-openhands}
\small
\begin{tabular}{@{}llp{8.5cm}@{}}
\toprule
\textbf{\#} & \textbf{Tool Name} & \textbf{Description} \\
\midrule
1  & \texttt{execute\_bash}          & Execute a bash command in a persistent shell session. Supports chaining, background processes, interactive input, and configurable timeouts. \\
2  & \texttt{browser}                & Interact with a web page via 15 browser actions: navigate, click, fill, scroll, hover, press, select option, double-click, focus, clear, drag-and-drop, upload file, go back/forward, and wait. \\
3  & \texttt{str\_replace\_editor}   & View, create, and edit files. Commands: \texttt{view}, \texttt{create}, \texttt{str\_replace}, \texttt{insert}, \texttt{undo\_edit}. Requires exact string matching for replacements. \\
4  & \texttt{execute\_ipython\_cell} & Run Python code in an IPython environment with persistent variables and \texttt{\%pip} magic support. \\
5  & \texttt{think}                  & Log a reasoning step for planning and brainstorming. Does not execute code or produce side effects. \\
6  & \texttt{finish}                 & Signal task completion with a summary message; ends the interaction. \\
7  & \texttt{task\_tracker}          & Structured task management. Commands: \texttt{view} (show task list) and \texttt{plan} (create/update tasks with status tracking). \\
\bottomrule
\end{tabular}
\end{subtable}

\vspace{1em}

\begin{subtable}{\textwidth}
\centering
\caption{Supabase Studio MCP tools (11).}
\label{tab:tools-supabase}
\small
\begin{tabular}{@{}llp{8.5cm}@{}}
\toprule
\textbf{\#} & \textbf{Tool Name} & \textbf{Description} \\
\midrule
1  & \texttt{execute\_sql}                & Execute a raw SQL query against the Postgres database. Intended for DML and one-off statements. \\
2  & \texttt{apply\_migration}            & Apply a DDL migration with version tracking. Requires a snake\_case name and a SQL query. \\
3  & \texttt{list\_tables}                & List all tables in one or more schemas (defaults to \texttt{public}). \\
4  & \texttt{list\_extensions}            & List all installed Postgres extensions. \\
5  & \texttt{list\_migrations}            & List all previously applied migrations. \\
6  & \texttt{get\_logs}                   & Fetch logs by service type (api, postgres, auth, storage, realtime, edge-function, branch-action) from the last 24 hours. \\
7  & \texttt{get\_advisors}               & Retrieve security or performance advisory notices (e.g.\ missing RLS policies). \\
8  & \texttt{get\_project\_url}           & Return the API URL for the Supabase project. \\
9  & \texttt{get\_anon\_key}              & Return the anonymous API key for the project. \\
10 & \texttt{generate\_typescript\_types} & Generate TypeScript type definitions from the database schema. \\
11 & \texttt{search\_docs}               & Search the Supabase documentation via a GraphQL query. Returns guides, CLI references, API references, and troubleshooting content. \\
\bottomrule
\end{tabular}
\end{subtable}

\vspace{1em}

\begin{subtable}{\textwidth}
\centering
\caption{Tavily MCP tools (4).}
\label{tab:tools-tavily}
\small
\begin{tabular}{@{}llp{8.5cm}@{}}
\toprule
\textbf{\#} & \textbf{Tool Name} & \textbf{Description} \\
\midrule
1  & \texttt{tavily-search}  & Real-time web search with customizable depth, domain filtering, time-range filtering, and support for general and news topics. Returns titles, URLs, content snippets, and optional images. \\
2  & \texttt{tavily-extract} & Extract and process content from specified URLs with basic or advanced parsing modes, including tables and embedded content. \\
3  & \texttt{tavily-crawl}   & Systematically crawl websites starting from a base URL with configurable depth, breadth limits, domain filtering, and path pattern matching. \\
4  & \texttt{tavily-map}     & Create structured site maps by analyzing website structure and navigation paths. Supports configurable depth, domain restrictions, and category filtering. \\
\bottomrule
\end{tabular}
\end{subtable}

\end{table*}

\clearpage
\section{Generation and Evaluation Hyper-Parameters}
\label{hyperparams}
\begin{table*}[!h]
\caption{Generation LLM hyper-parameters for all 16 models.}
\label{tab:generation_hparams_ranked}
\small
\begin{tabular}{@{}p{6.0cm} c c c c@{}}
\toprule
\textbf{Model ID} & \textbf{Temp} & \textbf{Top-p} & \textbf{Reasoning} & \textbf{Vision Disabled} \\
\midrule
\texttt{openai/gpt-5.3-codex-api-preview} & n/a & n/a & xhigh & - \\
\texttt{anthropic/claude-opus-4-6} & 1.0 & 1 & - & - \\
\texttt{openai/gpt-5.2} & n/a & 1 & xhigh & - \\
\texttt{anthropic/claude-sonnet-4-6} & 1.0 & 1 & - & - \\
\texttt{anthropic/claude-opus-4-6-thinking} & 1.0 & provider default & - & - \\
\texttt{openai/gpt-5.2-codex} & n/a & 1 & high & - \\
\texttt{google/gemini-3.1-pro-preview} & 1.0 & 1 & high & - \\
\texttt{zai/glm-5} & 1.0 & 0.95 & - & $\checkmark$ \\
\texttt{google/gemini-3-flash-preview} & 1.0 & provider default & high & - \\
\texttt{alibaba/qwen3.5-plus-thinking} & 0.6 & provider default & - & - \\
\texttt{kimi/kimi-k2.5-thinking} & 1.0 & 0.95 & - & - \\
\texttt{minimax/MiniMax-M2.5} & 1.0 & 0.95 & - & $\checkmark$ \\
\texttt{openai/gpt-5-mini-2025-08-07} & n/a & n/a & high & - \\
\texttt{anthropic/claude-haiku-4-5-20251001-thinking} & 1.0 & 1 & - & - \\
\texttt{deepseek-v3p2-thinking (via Fireworks)} & 1.0 & provider default & high & $\checkmark$ \\
\texttt{grok/grok-4-1-fast-reasoning} & 0.7 & 0.95 & - & - \\
\bottomrule
\end{tabular}
\end{table*}

All generation runs used the same API retry policy (maximum 20 retries with 15--120s backoff). Provider configuration (temperature, vision, reasoning effort) was chosen to elicit the best possible performance from each model.

\begin{table*}[!h]
\caption{UI evaluation hyper-parameters for the fixed evaluator used in reported runs.}
\label{tab:ui_eval_hparams_sonnet45}
\small
\begin{tabular}{@{}p{6.0cm} c c c c c@{}}
\toprule
\textbf{Parameter} & \multicolumn{5}{c}{\textbf{Value}} \\
\midrule
Evaluator model & \multicolumn{5}{c}{Claude Sonnet 4.5} \\
LLM adapter in Browser Use & \multicolumn{5}{c}{\texttt{ChatAnthropic(model)}} \\
Model sampling override & \multicolumn{5}{c}{$\checkmark$ (none set in evaluator code)} \\
Browser mode & \multicolumn{5}{c}{headless} \\
Window/viewport & \multicolumn{5}{c}{1920$\times$1200} \\
Agent flags & \multicolumn{5}{c}{\texttt{use\_vision=True}, \texttt{use\_thinking=True}, \texttt{calculate\_cost=True}} \\
LLM timeout (per call) & \multicolumn{5}{c}{90s} \\
Step timeout & \multicolumn{5}{c}{1200s} \\
Max agent steps (per workflow) & \multicolumn{5}{c}{100} \\
App startup timeout & \multicolumn{5}{c}{180s} \\
Per-app task timeout & \multicolumn{5}{c}{7200s} \\
Workflow pass threshold & \multicolumn{5}{c}{$\geq$90\% substeps passed} \\
\bottomrule
\end{tabular}
\end{table*}

\clearpage
\section{Generation System Prompt}
\label{sec:generation-prompt}
This was the prompt passed into the OpenHands harness to produce the application. Placeholders are replaced at runtime.

\begin{lstlisting}
You are a great software engineer developing new web applications.
You will be given a specification sheet for a web application meant to be used on a computer browser (usual computer screen, do not use mobile screen sizes). Your task is to build the application.
It is crucial your code works and follows the specification and deployment instructions given below. Focus hard when writing the code.
You CANNOT interact with the user at any point, you are on your own. Everything that is asked of you must be done in the code before you submit your work.
You CANNOT expect the user to perform any actions during or after the code is written.
DO NOT stop working until you have a fully working application that you have tested.
IT IS CRITICAL THAT YOU MAKE SURE TO TEST THE APPLICATION BEFORE SUBMITTING YOUR WORK, INCLUDING MAKING SURE YOU BUILD AND START THE DOCKER COMPOSE APPLICATION. DO NOT SUBMIT APPLICATIONS THAT DO NOT RUN. ALWAYS ASSUME ANY ERRORS IN BUILDING OR TESTING THE APPLICATION ARE YOUR FAULT AND YOUR RESPONSIBILITY TO FIX.

<TECH_STACK_GUIDELINES>
Tech stack guidelines:
Always use **React with Vite** as the framework for the frontend.
Do not use Next.js or any other framework unless explicitly requested.

Project scaffold guidelines:
1. The target directory already exists (e.g., "generated-app"). Create any necessary folders and files within this directory, **without** deleting any pre-existing files.
   - Example:
       cd generated-app
       yes | npm create vite@latest frontend -- --template react
   - **Do not overwrite `.env`** or any other files that the user has provided. .env is CRUCIAL for the development and testing of the application. You may copy some of the secrets to different places or append to this file, but always keep the original secrets intact.
2. Keep the project **minimal and lightweight**.
3. Treat **Supabase as the backend** (auth, database, storage, API) using the `@supabase/supabase-js` client (supabase details below).
4. Organize frontend code in a flat folder structure under `frontend/src/`:
    - `src/components/` -> reusable UI components
    - `src/pages/` -> page components
    - `src/hooks/` -> custom React hooks
    - `src/lib/` -> helper functions (e.g., Supabase client)
5. Use **React Router** for navigation if multiple pages are required.
6. Write clean, composable, and reusable code; avoid unnecessary boilerplate.
7. Follow Vite + React best practices:
    - Always use **ES modules** (`import`/`export`)
    - Lazy-load large components via `React.lazy` or dynamic imports
    - Do not modify Vite's default config unless necessary
8. Try to use Tailwind CSS v3. When scaffolding shadcn/ui, use `shadcn@2.3.0` for compatibility.

Backend Services (use only when necessary) for specific API endpoints:
  Use Express.js with Node.js for API endpoints that can't be handled by Supabase
  Examples of when a backend service is needed:
    Most likely: third-party API integrations that require server-side secrets: Stripe, etc.
    Less likely: complex business logic that requires server-side processing, but only if it is truly necessary.

  Keep REST API simple: use /api/[resource] structure
  Always validate inputs and handle errors gracefully
  Refer to the deployment setup for details on how to set it up and include it in the application
  Remember: most functionality should be handled by Supabase - only create backend services for truly necessary cases
</TECH_STACK_GUIDELINES>

<UI_UX_DESIGN_GUIDELINES>
UI/UX Design Guidelines:
  You must create a modern, professional interface. Make sure your nice design is actually rendered in the application.
  Follow these principles to create high-quality, professional interfaces:

  Visual Design:
    - Use consistent design system with unified color tokens, typography, spacing, and components
    - Limit typography to 4-5 font sizes and weights for visual hierarchy
    - Stick to 1 neutral base color (e.g., zinc) and up to 2 accent colors maximum
    - Always use multiples of 4 for padding and margins to maintain visual rhythm

  Component Architecture:
    - Build modular, reusable components to avoid duplication
    - Factor repeated UI patterns into shared components
    - Keep components small and focused, avoiding unnecessary complexity

  User Experience:
    - Use fixed height containers with internal scrolling for long content streams
    - Implement skeleton placeholders or `animate-pulse` for loading states
    - Indicate clickability with hover transitions (`hover:bg-*`, `hover:shadow-md`)
    - Provide clear visual feedback for user interactions and state changes
    - Ensure responsive design that works across different screen sizes

CRITICAL: The design system is everything. You should never write custom styles in components, you should always use the design system and customize it and the UI components (including shadcn components) to make them look beautiful with the correct variants. You never use classes like text-white, bg-white, etc. You always use the design system tokens.

- Maximize reusability of components.
- Leverage the index.css and tailwind.config.ts files to create a consistent design system that can be reused across the app instead of custom styles everywhere.
- Create variants in the components you'll use. Shadcn components are made to be customized!
- You review and customize the shadcn components to make them look beautiful with the correct variants.
- CRITICAL: USE SEMANTIC TOKENS FOR COLORS, GRADIENTS, FONTS, ETC. It's important you follow best practices. DO NOT use direct colors like text-white, text-black, bg-white, bg-black, etc. Everything must be themed via the design system defined in the index.css and tailwind.config.ts files!
- Always consider the design system when making changes.
- Pay attention to contrast, color, and typography.
- Always generate responsive designs.
- Beautiful designs are your top priority, so make sure to edit the index.css and tailwind.config.ts files as often as necessary to avoid boring designs and leverage colors and animations.
- Pay attention to dark vs light mode styles of components. You should avoid making mistakes like having white text on white background and vice versa. You should make sure to use the correct styles for each mode.
</UI_UX_DESIGN_GUIDELINES>

<SUPABASE_INTEGRATION>
Supabase Integration:
  You must use Supabase for database, authentication, and storage.
  The Supabase service is already running and configured. You SHOULD NOT reproduce any of the Supabase services locally.
  Focus on using the MCP tools and the SDK to interact with the Supabase service as described below. You cannot use the default CLI as if you were accessing Supabase Cloud.
  The keys and URLs you will need are in the .env file: `SUPABASE_PROJECT_URL`, `SUPABASE_ANON_KEY`, `SUPABASE_SERVICE_ROLE_KEY`.

  Database:
    - Use the MCP tools for all database operations (you have access to a self-hosted Supabase database)
    - The only way to interact with the database is through the MCP tools available to you
    - Do not create separate database services in Docker
    - When calling MCP SQL tools, pass statements via the `query` argument (both `execute_sql` and `apply_migration` expect it)
    - Respect Row Level Security (RLS) and database constraints. Define explicit foreign keys and unique indexes for all relationships, and ensure all inserts/updates satisfy referential integrity to avoid "missing relation" and "duplicate key" errors.
    - Capture schema changes as real migrations: prefer the `apply_migration` tool so Supabase records the version, and check the generated SQL into `supabase/migrations/`. If you must fall back to `execute_sql`, send exactly one statement per call so the tool can parse the result reliably.
    - For admin/seed operations that require bypassing RLS, use the service role key; for user-facing flows, use the anon key and create appropriate RLS policies. Never disable RLS.
    - Apply schema migrations and policies via the MCP SQL tools before running the app.

  Seeding & Test Data:
    - Provide a minimal seed routine that creates the records required by the spec and tests (e.g., at least one event and organizer with valid relations).
    - Make seeding idempotent: upsert by natural keys to avoid duplicate key violations.
    - Document how to run the seed and run it before UI testing.

  Authentication & Storage:
    - Use Supabase SDK for authentication and file storage
    - Use MCP tools for server-side database operations and setup.
    - The Studio MCP server does not expose auth-user helpers. For seeding or admin automation, call Supabase Auth directly using the service role key -- either via `supabase.auth.admin.*` in the SDK or by POSTing to `/auth/v1/admin/users` through Kong. Document any seeded credentials in the README.
    - Implement proper auth flows (login, signup, logout, protected routes), if it is required by the spec.
    - Handle authentication state management properly
    - For the authentication, the email verification is disabled, as well as the third-party authentication like Google, Apple, etc. Do not try to use them.
    - If there is an authentication flow, strongly test it, as it is the entry point of the application and if it doesn't work, the application won't work.
    - Use Supabase storage for file uploads when needed

  Additional Services:
    - You only can use Supabase authentication, database, and storage.
    - You DO NOT have access to edge functions, neither with MCP nor directly with SDK/CLI.
    - Any other services you may need must be written in the backend service, only if it is crucial.

  Configuration:
    - DO NOT hardcode any API keys, secrets, or URLs in the docker-compose.yml file
    - All sensitive configuration must be in a .env file. The .env file already contains the necessary URL and API keys for you to use the Supabase service.
    - Reference Supabase documentation if needed: https://supabase.com/docs and https://supabase.com/docs/reference.
    - Keep in mind the user will not be able to access the Supabase service once you have submitted the app.
</SUPABASE_INTEGRATION>

<EMAIL_SERVICES>
Email Services:
  - You have access to a MailHog SMTP server, that will keep running during both development and testing.
  - You can use it to create email sending workflows in your application. The server can handle sending and receiving from and to any email address.
  - The application should send emails via SMTP (config in .env: MAILHOG_SMTP_HOST, MAILHOG_SMTP_PORT)
  - Use MailHog without SMTP auth; omit the auth block unless both SMTP_USER and SMTP_PASS are provided.
  - You can access the website at the MAILHOG_WEB_URL (in .env) to see the emails that are sent.
  - Use appropriate libraries: nodemailer for Node.js, smtplib for Python, etc.
  - ONLY use the MailHog server to send emails, do not try anything else.
</EMAIL_SERVICES>

<STRIPE_SERVICES>
Stripe Services:
  - Do NOT rely on MCP tools for Stripe; use the official Stripe SDK only.
  - Read `STRIPE_SECRET_KEY` and `STRIPE_ACCOUNT_ID` and `STRIPE_PUBLISHABLE_KEY` from environment (.env) and use them to initialize the SDK.
  - You MUST use the Stripe account ID to make sure your work is isolated to that account.
  - Keep the Stripe integration simple and minimal: just a Stripe-hosted Checkout page for payment/subscription with success URL and cancel URL.
  - Never hardcode secrets. All keys/IDs must come from environment variables.
  - Never send raw credit card numbers directly to Stripe; this is unsafe and requires strict PCI compliance.
  - All card entry must happen on the Stripe-hosted Checkout page.
  - The currency is USD, always use it for all payments, products, and prices.
  - For testing, use credit card number 4242 4242 4242 4242 with any CVC and valid future expiration date.

### Required Stripe + Supabase Integration Flow:

**Steps:**
1. **Create Checkout Session** (server-side API route)
   - Link Stripe customer to Supabase user via metadata
   - Create payment record in Supabase with 'pending' status
   - Set success_url and cancel_url pointing to your localhost app
   - Return session URL to client

2. **Success Page Verification** (CRITICAL!)
   - User returns with session_id parameter
   - Server-side: Retrieve session from Stripe API to verify payment status
   - Update Supabase tables based on verified payment status
   - Never trust client-side data for payment fulfillment

**Implementation:**
- Create API routes: `/api/create-checkout` and `/api/verify-payment`
- Always verify payments server-side using `stripe.checkout.sessions.retrieve(sessionId)`
- Update user profiles and payment records in Supabase after verification
- Works perfectly with localhost app + online Stripe services

**IMPORTANT:** Use session status verification on success page instead of webhooks for development. This is reliable and works with localhost environments.
</STRIPE_SERVICES>

<DEPLOYMENT_INSTRUCTIONS>
# Deployment Instructions
Docker and Docker Compose are already installed in the environment. Do not try to install them or change the installed versions.
The Docker daemon should also be running in the background. Only try to start it if you encounter an error such as "Is the docker daemon running?".
When you start the app, it should be available at the APP_PUBLIC_URL and APP_PUBLIC_PORT, using localhost.
The different services (like Supabase) you have access to are using a different host, because they are running in a different place than your development environment. Keep it that way for those services, but for your application, use localhost.
Inside the frontend, you should not hardcode any backend or Supabase URLs you query; instead, you should append them as variables to a .env file.

You are asked to make the application runnable with Docker Compose.
It must be runnable from the generated-app directory, i.e., {{workspace_path}}/generated-app/ in your case.
The project structure should be:

```
{{workspace_path}}/
  .browser_screenshots/ (DO NOT DELETE - used for testing)
  .downloads/ (DO NOT DELETE - used for testing)
  generated-app/
    .env (DO NOT DELETE OR OVERWRITE - this file is provided before you start working on the application. It contains provided environment variables you will need for your application. Do not overwrite it. You may need to copy the values to other .env file).
    docker-compose.yml
    frontend/
      .env (if necessary)
      .dockerignore
      Dockerfile.frontend # Dockerfile for the frontend
      src/ # folder created by the Vite CLI, containing the frontend code
      ...other folders/files...
    backend/ # Only create a backend if necessary
      .env (if necessary)
      .dockerignore
      Dockerfile.backend
    ...other folders/files...
```

The backend folder is optional - you only need to create it if necessary.

If you use CLI tools to initialize the app, never use options that can delete existing files you are provided with, especially the .env file.
This .env file contains crucial information for you to build the application and to test it.

## Docker Compose Setup
- Create a single `docker-compose.yml` file in the project root ({{workspace_path}}/generated-app/)
- Name the frontend service exactly `{project_id}` (this service will be exposed)
- Do NOT use `networks` configuration - only fill in the `services` section
- Do NOT use `container_name` fields
- Remember that at application test time, there will be no cached Docker images, so you must make sure the build command works.
- Always run the container by building and then previewing it (e.g. `npm run build` followed by `npm run preview`, `npm run serve`, or `npm start`); avoid `npm run dev`.
- In case you need to save data outside of what Supabase already provides, DO NOT rely on Docker Compose creating persistent volumes for your app's data, because in the testing environment those volumes might not be preserved. Instead, any data your app needs to persist should be stored in normal files inside the container (or a path you control), so that the tests can access it reliably.
- Keep the docker-compose.yml and the Dockerfiles minimal, as simple as possible.
- This doesn't need to be production-ready; it just needs to run in development/preview mode. A simple Dockerfile is enough -- no need for a multi-stage or optimized production setup.
- DO NOT use the network_mode field in the docker-compose.yml file.

## Port Configuration
- The app landing page URL is defined by `APP_PUBLIC_URL` in the `.env`; treat it as the source of truth and never assume `http://localhost:4173`.
- Always map the host port from `APP_PUBLIC_PORT` to the frontend preview port `4173` in Docker Compose (for example, `ports: - "${APP_PUBLIC_PORT}:4173"`).
- Do not override `APP_PUBLIC_URL` or `APP_PUBLIC_PORT` inside your code; surface them through environment variables so both frontend and backend use the same public origin.
- Always import the port from the .env file, never hardcode it in the application code anywhere.
- Configure Vite to listen on all interfaces for both dev and preview (set `server.host`/`preview.host` to `0.0.0.0` or pass `--host 0.0.0.0`).
- If you add backend services that are internal-only, prefer `expose: - 8000` instead of publishing them to the host.

## Example of minimal docker-compose.yml and Dockerfile (they might need to be more complex depending on the project). These match all the requirements described above. You might need them to be more complicated depending on the project (especially if you need backend services), but always strive to keep it simple:
```dockerfile
FROM node:22-alpine
WORKDIR /app
COPY package*.json ./
RUN npm install
COPY . .
RUN npm run build
EXPOSE 4173
CMD ["npm", "run", "preview"]
```

```yaml
services:
  donation_manager:
    build:
      context: ./frontend
      dockerfile: Dockerfile.frontend
    ports:
      - "${APP_PUBLIC_PORT}:4173"  # APP_PUBLIC_PORT should always be imported from the .env file
    env_file:
      - .env

  backend: # Only use a backend service if necessary
    build:
      context: ./backend
      dockerfile: Dockerfile.backend
    env_file:
      - .env
    expose: # NEVER use ports: - 8000:8000 kind of configuration here, as it would publish the backend port outside of the Docker Compose network
      - 8000
```

Example of a minimal backend Dockerfile (if necessary):
```
FROM node:22-alpine
WORKDIR /app
COPY package*.json ./
RUN npm install
COPY . .
# Copy only what the backend needs, not the entire frontend
EXPOSE 8000
CMD ["node", "server.js"]
```

## Docker Ignore
Always include a `.dockerignore` file in any relevant folder (e.g. frontend/, backend/, etc.) to avoid copying unnecessary files into the container image.
At minimum, ignore build outputs and dependency folders such as: node_modules, dist, build, etc.
Example of a `.dockerignore` file:
```
node_modules
.next
dist
.git
```

## Development & Testing Workflow
1. **Initial build (and try before submitting)**: `docker compose -p {project_id} up -d --build`
2. **After code changes**: `docker compose -p {project_id} restart <service>`
3. **Only rebuild when**: dependencies or docker-compose.yml change
4. **Check the logs to make sure the application is running correctly**: Run `docker compose -p {project_id} logs --tail 200` (or wrap `logs -f` in a short `timeout`) so the command exits on its own.
5. **Test the app through the UI**: You must browse the running application and take screenshots (you MUST save them in `{{workspace_path}}/.browser_screenshots/`) to verify functionality
6. **Before submitting**: Run `docker compose -p {project_id} down -v`
7. **Report**: Report how testing went in your final thoughts before submitting your work. Mention what worked or did not work, and whether you followed the expected workflow. Report precisely any deployment errors or issues you encountered.

## Mandatory Testing
- Browse the running application and take screenshots to verify functionality
- Fix issues and restart services until you've verified everything works
- Save all screenshots in `{{workspace_path}}/.browser_screenshots/`

The application must be fully functional when run with the following command from {{workspace_path}}/generated-app/:
```bash
docker compose -p {project_id} up -d --build
```
</DEPLOYMENT_INSTRUCTIONS>

Absolutely No Mocks or Stubs (Hard Requirement):
  - Do not mock payments or return hardcoded success for checkout/subscription flows. A real Stripe test-mode call must occur.
  - Do not mock email sending; use SMTP with the env values and send test emails (MailHog will capture them in dev).
  - Avoid in-memory replacements for required services when the spec calls for them (e.g., do not replace Supabase/Stripe with in-memory data for subscriptions or billing flows).
  - Submissions that include phrases like "mock Stripe", "fake payment", or bypass external calls will be rejected.

<README_INSTRUCTIONS>
README instructions:
  You must create two READMEs for the application:
  First, a technical TECHNICAL_README.md at project root in {{workspace_path}}/generated-app/, at most one page long, with:
  1. General Implementation
      - App overview and key features
      - Tech stack summary
      - Architecture approach
  2. Code Structure
      - Frontend/backend directory layout
      - Key files and their purposes
      - Database schema

  Second, a user-friendly USER_README.md at project root in {{workspace_path}}/generated-app/, at most one page long, with:
  1. Navigation Guide to Test the Application
      - Authentication (admin or not) method if any
      - Key URLs and actions
      - If there is an authentication flow, provide default credentials to be able to test the application without recreating an account.

  Note: Exclude startup instructions - app runs via Docker commands in deployment docs
</README_INSTRUCTIONS>

<GENERAL_INSTRUCTIONS_AND_ADVICE>
General Instructions and Advice:
  - The folder you are working in is `{{workspace_path}}`. You cannot write outside of this folder. It also may contain the following files and folders, that you CANNOT delete in any circumstance (it would break the application):
  ```
  {{workspace_path}}/
    .browser_screenshots/
    .downloads/
    generated-app/
  ```
  - You must create your application in `{{workspace_path}}/generated-app/` directory. This is where your docker-compose.yml, .env file, and all application code should be placed.
  - The .browser_screenshots/ and .downloads/ folders at the `{{workspace_path}}` level must NEVER be deleted or moved - they are used for testing purposes.
  - If you have trouble loading data in the application, make sure to `docker compose -p {{project_id}} down -v` before running `docker compose -p {{project_id}} up -d` again. It will clear the volumes and start fresh.
  - Ensure proper error handling and user feedback for all operations.
  - Strictly restrict yourself to the services you have access to. Do not try to implement services that are not available to you. If a feature requires an additional service that you do not have access to, just ignore that feature, or worst case create a button that just shows an error "missing service".
  - If you encounter issues with the infrastructure you are running in, make sure to put them in your final thoughts before submitting your work. For example, mention any problems you encountered when trying to test the application, especially Docker-related issues. Make sure to report the commands that failed with the error message.
  - You must focus on doing the full task. Do not abandon until you have completed the task or try as much as possible to complete the task. If you encounter errors that you believe are not expected, do not give up, try to fix them or go around them.
  - You are on your own, you CANNOT interact with the user at any point. Do not expect them to take any actions after the code is written. The only thing the user will do is use the application and see if it is working as expected.
  - Before running any UI tests, run the project's seed routine to ensure required data exists and relationships are valid (avoid RLS and constraint errors during evaluation).
</GENERAL_INSTRUCTIONS_AND_ADVICE>

<TASK_INSTRUCTIONS>

Specification for the application. Follow the instructions below closely to build the application:
{{spec}}

</TASK_INSTRUCTIONS>
\end{lstlisting}

\section{Spec Example}
\label{sec:appendix-spec-example}
\begin{lstlisting}
Zeeter Social Network

Overview

Deliver a browser-based short-form social app where members publish text updates, follow users, and interact through likes, comments, and in-app notifications. Guests can browse public content but cannot perform member actions.

User Roles

- Guest: Browses public profiles and public posts.
- Member: Signs up, manages profile, creates/edits/deletes own posts, follows other members, likes and comments on posts, and views notifications.

Authentication

- Members sign up with username, email, and password.
- Members sign in with username or email plus password.
- Email verification is not required.
- Guests can browse public content without signing in.

Core Features

1. Accounts and Profiles

- Members can create accounts with username, email, and password.
- Members can set and update display name, bio, and profile image.
- Public profile pages show member identity fields and recent posts.

2. Posts

- Members can create posts from the home feed.
- Post content is limited to 280 characters.
- Members can edit and delete only their own posts.

3. Feed and Discovery

- Member home feed includes followed-member content and discovery content.
- Feed supports sorting by Newest and Trending.
- Feed supports search by keyword and hashtag.

4. Social Interactions

- Members can follow other members.
- Members can like posts.
- Members can comment on posts.

5. In-App Notifications

- Notifications page shows follow, like, and comment events relevant to the member.

Content and Constraints

- Text-only posting experience for this version (no media in posts).
- All key workflows must be executable in one browser session.

Primary User Flows

- Onboard and profile setup: sign up, complete profile fields, land on feed.
- Publish and manage posts: create a post, edit it, and delete it.
- Social engagement: follow, like, and comment across two member accounts.
- Guest restrictions: browse public content but block member-only actions.
- Discovery usage: sort feed and search by hashtag/keyword.

\end{lstlisting}

\section{UI Test Example}
\label{sec:appendix-ui-test-example}
\begin{lstlisting}
test_workflows:
  onboarding_and_profile_setup:
    id: id_0
    title: Onboarding and Profile Setup
    purpose: Verify a new member can sign up with username, email, and password, complete profile setup, and land on the home feed.
    num_substeps: 6
    steps:
      - 1. Navigate to the Zeeter app and open the sign-up page.
      - 2. Sign up with username 'zeeter_0', email 'zeeter_0@example.com', and password 'UserPass123!'.
      - 3. Complete profile setup with display name 'Zeeter Zero' and bio 'First Zeeter profile setup'.
      - 4. Upload a valid profile image and save profile changes.
      - 5. Verify the app navigates to the signed-in home feed.
      - 6. Open the public profile for 'zeeter_0' and verify display name, bio, and profile image are visible.

  update_profile_fields:
    id: id_1
    title: Update Profile Fields
    purpose: Verify a member can update display name, bio, and profile image after account creation.
    num_substeps: 6
    steps:
      - 1. Sign up with username 'zeeter_1', email 'zeeter_1@example.com', and password 'UserPass123!'.
      - 2. Open the profile settings page.
      - 3. Change display name to 'Zeeter One Updated' and update bio to 'Updated Zeeter profile bio'.
      - 4. Upload or replace the profile image.
      - 5. Save changes and verify an in-app success confirmation appears.
      - 6. Open the public profile for 'zeeter_1' and verify updated fields are displayed.

  create_post_from_feed:
    id: id_2
    title: Create Post from Feed
    purpose: Verify a member can publish a post from the feed and the post appears in profile history.
    num_substeps: 6
    steps:
      - 1. Sign up with username 'zeeter_2', email 'zeeter_2@example.com', and password 'UserPass123!'.
      - 2. On the home feed, locate the post composer.
      - 3. Enter post text 'Gardening update #zeeterpost'.
      - 4. Publish the post.
      - 5. Verify the post appears in the feed.
      - 6. Open the public profile for 'zeeter_2' and verify the same post appears in recent posts.

  edit_and_delete_own_post:
    id: id_3
    title: Edit and Delete Own Post
    purpose: Verify a member can edit and delete only their own post.
    num_substeps: 8
    steps:
      - 1. Sign up with username 'zeeter_3', email 'zeeter_3@example.com', and password 'UserPass123!'.
      - 2. Create a post with content 'Original Zeeter message'.
      - 3. Open edit for that post, replace content with 'Updated Zeeter message', and save.
      - 4. Verify the post now shows only 'Updated Zeeter message'.
      - 5. Delete the same post and confirm deletion.
      - 6. Verify the deleted post is no longer visible in the home feed.
      - 7. Open the public profile for 'zeeter_3'.
      - 8. Verify the deleted post is not listed in recent posts.

  visitor_browsing_and_restrictions:
    id: id_4
    title: Visitor Browsing and Restrictions
    purpose: Verify guests can browse public content but cannot perform member-only actions.
    num_substeps: 9
    steps:
      - 1. Sign up with username 'zeeter_4', email 'zeeter_4@example.com', and password 'UserPass123!'.
      - 2. Create a post with content 'Public post for guest access check'.
      - 3. Open the public profile for 'zeeter_4' and then sign out.
      - 4. As a guest, navigate back to 'zeeter_4' public profile.
      - 5. Verify the public post is visible.
      - 6. Attempt to like the post and verify the app blocks the action or prompts sign-in.
      - 7. Attempt to comment on the post and verify the app blocks the action or prompts sign-in.
      - 8. Navigate to the home feed and verify post creation is unavailable to guests.
      - 9. Attempt to follow 'zeeter_4' as a guest and verify the app blocks the action or prompts sign-in.

  follow_member_and_feed_update:
    id: id_5
    title: Follow Member and Feed Update
    purpose: Verify following a member updates the follower feed and creates a notification for the followed member.
    num_substeps: 11
    steps:
      - 1. Sign up with username 'zeeter_5_a', email 'zeeter_5_a@example.com', and password 'UserPass123!'.
      - 2. Create a post with content 'Follow workflow seed post #followseed'.
      - 3. Sign out from 'zeeter_5_a'.
      - 4. Sign up with username 'zeeter_5_b', email 'zeeter_5_b@example.com', and password 'UserPass123!'.
      - 5. Navigate to the public profile for 'zeeter_5_a'.
      - 6. Follow 'zeeter_5_a' and verify follow state is active.
      - 7. Open the home feed and verify the seed post from 'zeeter_5_a' appears.
      - 8. Sign out from 'zeeter_5_b'.
      - 9. Sign in with username 'zeeter_5_a' (email 'zeeter_5_a@example.com') and password 'UserPass123!'.
      - 10. Open notifications page.
      - 11. Verify a new-follower notification referencing 'zeeter_5_b' is present.

  like_comment_and_notifications:
    id: id_6
    title: Like, Comment, and Notifications
    purpose: Verify likes and comments generate expected UI state changes and notifications.
    num_substeps: 11
    steps:
      - 1. Sign up with username 'zeeter_6_a', email 'zeeter_6_a@example.com', and password 'UserPass123!'.
      - 2. Create a post with content 'Interaction workflow seed post #interactionseed'.
      - 3. Sign out from 'zeeter_6_a'.
      - 4. Sign up with username 'zeeter_6_b', email 'zeeter_6_b@example.com', and password 'UserPass123!'.
      - 5. Open the public profile for 'zeeter_6_a' and locate the seed post.
      - 6. Like the post and verify the like state updates.
      - 7. Submit comment 'Great update from zeeter_6_b'.
      - 8. Verify the comment appears under the post.
      - 9. Sign out from 'zeeter_6_b'.
      - 10. Sign in with username 'zeeter_6_a' (email 'zeeter_6_a@example.com') and password 'UserPass123!'.
      - 11. Verify notifications include both the like and comment from 'zeeter_6_b'.

  feed_sorting_newest_and_trending:
    id: id_7
    title: Feed Sorting by Newest and Trending
    purpose: Verify feed sorting supports Newest ordering and Trending engagement-based discovery.
    num_substeps: 10
    steps:
      - 1. Sign up with username 'zeeter_7_a', email 'zeeter_7_a@example.com', and password 'UserPass123!'.
      - 2. Create post P1 with content 'Trending seed post #zeetertrend'.
      - 3. Create post P2 with content 'Newest seed post #zeetertrend'.
      - 4. Sign out from 'zeeter_7_a'.
      - 5. Sign up with username 'zeeter_7_b', email 'zeeter_7_b@example.com', and password 'UserPass123!'.
      - 6. Navigate to 'zeeter_7_a' profile and like P1.
      - 7. Sign out from 'zeeter_7_b' and sign in with username 'zeeter_7_a' (email 'zeeter_7_a@example.com') and password 'UserPass123!'.
      - 8. Set feed sorting to Newest and verify P2 appears above P1.
      - 9. Set feed sorting to Trending.
      - 10. Verify P1 ranks above P2 in Trending based on the like received from 'zeeter_7_b'.

  search_by_keyword_and_hashtag:
    id: id_8
    title: Search by Keyword and Hashtag
    purpose: Verify search supports both keyword and hashtag queries.
    num_substeps: 7
    steps:
      - 1. Sign up with username 'zeeter_8', email 'zeeter_8@example.com', and password 'UserPass123!'.
      - 2. Create a post with content 'Gardening updates #urbanfarm'.
      - 3. Create a second post with content 'Book club planning tonight'.
      - 4. Search for hashtag '#urbanfarm'.
      - 5. Verify the gardening post appears and the 'Book club' post is not shown in hashtag results.
      - 6. Search for keyword 'Book club'.
      - 7. Verify the 'Book club planning tonight' post appears in results.

  post_character_limit_validation:
    id: id_9
    title: Post Character Limit Validation
    purpose: Verify post creation enforces the 280-character limit.
    num_substeps: 6
    steps:
      - 1. Sign up with username 'zeeter_9', email 'zeeter_9@example.com', and password 'UserPass123!'.
      - 2. Attempt to publish a post with text 'A' repeated 281 times.
      - 3. Verify publishing is blocked and an in-app validation message indicates the content exceeds the limit.
      - 4. Enter a post with text 'B' repeated 280 times.
      - 5. Publish the 280-character post.
      - 6. Verify the post is created and visible in the feed.

\end{lstlisting}

\end{document}